\documentclass[aps,pre,twocolumn,superscriptaddress,showpacs,floatfix]{revtex4-1}
\usepackage{dcolumn}
\usepackage{amsmath}
\usepackage{amssymb}
\usepackage{graphicx}
\usepackage{bm}
\usepackage[T1]{fontenc}
\usepackage{color}
\usepackage{url}
\usepackage[bf]{subfigure}
\usepackage{rotating}
\usepackage{times}

\usepackage{scalefnt}
\usepackage{multirow}
\usepackage{threeparttable}
\usepackage{float}

\begin{document}

\title{Unifying Markov Chain Approach for Disease and Rumor Spreading in Complex Networks}

\author{Guilherme Ferraz de Arruda}
\affiliation{Departamento de Matem\'{a}tica Aplicada e Estat\'{i}stica, Instituto de Ci\^{e}ncias Matem\'{a}ticas e de Computa\c{c}\~{a}o,
Universidade de S\~{a}o Paulo - Campus de S\~{a}o Carlos, Caixa Postal 668, 13560-970 S\~{a}o Carlos, SP, Brazil.}

\author{Francisco A. Rodrigues}
\email{francisco@icmc.usp.br}
\affiliation{Departamento de Matem\'{a}tica Aplicada e Estat\'{i}stica, Instituto de Ci\^{e}ncias Matem\'{a}ticas e de Computa\c{c}\~{a}o,
Universidade de S\~{a}o Paulo - Campus de S\~{a}o Carlos, Caixa Postal 668, 13560-970 S\~{a}o Carlos, SP, Brazil.}

\author{Pablo Mart\'in Rodr\'iguez}
\affiliation{Departamento de Matem\'{a}tica Aplicada e Estat\'{i}stica, Instituto de Ci\^{e}ncias Matem\'{a}ticas e de Computa\c{c}\~{a}o,
Universidade de S\~{a}o Paulo - Campus de S\~{a}o Carlos, Caixa Postal 668,
13560-970 S\~{a}o Carlos, SP, Brazil.}

\author{Emanuele Cozzo}
\affiliation{Institute for Biocomputation and Physics of Complex Systems (BIFI), University of Zaragoza, 50018 Zaragoza, Spain}
\affiliation{Department of Theoretical Physics, University of Zaragoza, 50018 Zaragoza, Spain}

\author{Yamir Moreno}
\email{yamir.moreno@gmail.com}
\affiliation{Institute for Biocomputation and Physics of Complex Systems (BIFI), University of Zaragoza, 50018 Zaragoza, Spain}
\affiliation{Department of Theoretical Physics, University of Zaragoza, 50018 Zaragoza, Spain}
\affiliation{Complex Networks and Systems Lagrange Lab, Institute for Scientific Interchange, Turin, Italy}

\begin{abstract}
Spreading processes are ubiquitous in natural and artificial systems. They can be studied via a plethora of models, depending on the specific details of the phenomena under study. Disease contagion and rumor spreading are among the most important of these processes due to their practical relevance. However, despite the similarities between them, current models address both spreading dynamics separately. In this paper, we propose a general information spreading model that is based on discrete time Markov chains. The model includes all the transitions that are plausible for both a disease contagion process and rumor propagation. We show that our model not only covers the traditional spreading schemes, but that it also contains some features relevant in social dynamics, such as apathy, forgetting, and lost/recovering of interest. The model is evaluated analytically to obtain the spreading thresholds and the early time dynamical behavior for the contact and reactive processes in several scenarios. Comparison with Monte Carlo simulations shows that the Markov chain formalism is highly accurate while it excels in computational efficiency. We round off our work by showing how the proposed framework can be applied to the study of spreading processes occurring on social networks.
\end{abstract}

\pacs{89.75.Hc,89.75.-k,89.75.Kd}

\maketitle

\section{Introduction}

The spreading of diseases and information are two processes intimately linked that have been the subject of intense study since long time ago. These contagion phenomena are pervasive in nature, society and engineering~\cite{Satorras015}. As a result, we nowadays are capable to describe and understand many aspects of the mechanisms behind the propagation of pathogens among humans and other species, of digital viruses and malware through the Internet and diverse socio-technical systems, and of rumors among individuals, to mention a few examples~\cite{Satorras015}. Moreover, with the advent of modern communication technologies and transportation means and the increasing availability of data (often in real time or with a highly detailed time resolution), previous theoretical-only models are being fed with data, making data-driven simulations an effective tool for decision-making and for the design of efficient viral algorithms in the case of information dissemination.

The early models dealing with epidemic and rumor spreading considered only homogeneous mixing, in which the probability that a given node interacts with any other node in the network is the same for all nodes ~\cite{Keeling05}. Although the homogeneous approximation facilitates the theoretical analyses of contagion processes, this approach turned out to be too simple as to capture the architecture of real-world complex systems, whose structure is largely heterogeneous \cite{Boccaletti06}. Indeed, the availability of data about the topology of real systems spurred the development of network models~\cite{Barabasi99}, which in turn led to the inclusion of network's interaction patterns into disease spreading models~\cite{Pastor-Satorras2001}. 

One of the first and most used methods to study disease (and rumor) propagation is the so-called degree-based mean-field (DMF) approximation~\cite{Pastor-Satorras2001}. This approach groups vertices into classes and assumes that all nodes with the same degree are equivalent from a dynamical point of view. However, the DMF does not provide information about the probability of individual nodes. Recently, a formalism based on probabilistic discrete-time Markov chains was introduced to generalize existing MF approximations~\cite{Gomez2010}. Differently from the MF approximation, discrete-time Markov chains enable the description of individual node dynamics as well as the determination of the macroscopic critical properties and the whole phase diagram~\cite{Gomez2010}. 

Epidemic and rumor spreading processes are similar in many aspects. Indeed, the creation mechanism is the same in classical models: with a given (spreading) probability, a disease (rumor) is transmited to any of the neighbors of an infected (spreader) individual. On the contrary, the annihilation mechanisms are completely different by the very nature of the processes being studied. In disease contagion, spreaders die out because they recover from the infection with a given probability. This is independent of their neighbors' dynamical states and of any interaction. Rumor-like processes are fundamentally different: in traditional models, rumors decay as a result of the interactions between spreaders and other individuals that already know the rumor $-$ no matter whether they are actually propagating it (spreaders) or have already stopped (stiflers). Despite the differences between these two contagion processes, they have been studied using the same methodological approaches, albeit independently~\cite{Satorras015}. Surprisingly, the phenomenological similarities between them have not been fully exploited and there is not a general framework that allows studying both processes under the same formulation. Admittedly, rumor-like dynamics on complex topologies have been studied only at the mean-field level, and approaches like the discrete-time Markov chains are not available yet. 

Here, we fill the existent gap and propose a general information-spreading model that captures the dynamical behavior of epidemic and rumor spreading processes. Our model includes, as particular cases, not only the traditional models of rumor and epidemic spreading, but also other mechanisms such as apathy~\cite{Borge-Holthoefer2012}, forgetting~\cite{Kawachi200855}, lost of interest~\cite{Nekovee2007, Kawachi200855} and a new mechanism proposed here that characterizes cases in which the interest in the rumor can be recovered once lost. We thoroughly study, both analytically and numerically, the dynamics of the model in random synthetic graphs as well as in real social networks. Moreover, we analyze several plausible scenarios and obtain the corresponding critical spreading probabilities for the contact and reactive limiting cases, in addition to analyze the early time dynamics of the spreading process. We also perform extensive numerical simulations and show that the discrete Markov chain approach developed here is highly accurate, both at the micro and the macroscopic levels. Finally, we discuss potential applications of this framework in the context of social contagion and the design of new viral algorithms for efficient information dissemination.

The rest of the manuscript is organized as follows. In the next section, we present some related works summarizing the main previous results and approaches that have been adopted so far. Section~\ref{sec:model} presents our model. Its early time behavior and the steady state analysis are presented in Section~\ref{sec:Analytical}. Section \ref{sec:special} analyzes several special cases before turning our attention to the numerical analysis of the model, which is performed in Section~\ref{sec:Numerical}. Finally, applications are discussed in Section~\ref{sec:social}, whereas our conclusions are presented in Section~\ref{Sec:conclusion}.


\begin{table*}[!t] 
\centering
\begin{threeparttable}[b]
\caption{A brief literature review: A summary of previous epidemic and rumor spreading models. The states are (i) susceptible or ignorant ($X$), (ii) infected or spreader ($Y$) and (iii) recovered or stifler ($Z$). }
\begin{tabular}{|c|l|c|c|c|c|c|}
\hline
& Model & Interactions & Threshold & Networks & Comments & References \\
\hline
\hline
\multirow{9}{*}{\vspace{-1cm}\hspace{0.2cm}\begin{rotate}{90}
\hbox{Epidemic}
\end{rotate}\hspace{0.2cm}} 
& SI   & $Y+X \xrightarrow{\lambda} Y+Y$ & \hspace{.7cm} -- \hspace{.7cm} & Yes & Only two fixed points. & \cite{Barrat08:book}, \cite{Newman010:book}\\
\cline{2-7}
& \multirow{2}{*}{SIR}  & $Y+X \xrightarrow{\lambda} Y+Y$ & \multirow{2}{*}{$\lambda > \frac{\delta}{\Lambda_{max}}$}~\tnote{1} & \multirow{2}{*}{Yes} & Absorbing state, $R$; & \multirow{2}{*}{\cite{Barrat08:book}, \cite{Newman010:book}}\\
&      & $Y \xrightarrow{\delta} Z$      & &  & Presence of influential spreaders. & \\
\cline{2-7}
& \multirow{2}{*}{SIS}  & $Y+X \xrightarrow{\lambda} Y+Y$ & \multirow{2}{*}{$\lambda > \frac{\delta}{\Lambda_{max}}$}\tnote{2} & \multirow{2}{*}{Yes} & Presents an active steady state; & \cite{Pastor-Satorras2001}, \cite{Boguna2002}, \cite{Moreno2002},  \\
&      & $Y \xrightarrow{\delta} X$      & &  & Discrete and continuous time. & \cite{Boguna2003}, \cite{Moreno2003}, \cite{Barrat08:book}, \cite{Gomez-Gardenes2008},\\
& & & & & &  \cite{Mieghem09}, \cite{Newman010:book}, \cite{Gomez2010}, \cite{Boguna2013} \\
\cline{2-7}
& \multirow{3}{*}{SIRS} & $Y+X \xrightarrow{\lambda} Y+Y$ & \multirow{3}{*}{$\lambda > \delta \frac{\langle k \rangle}{\langle k^2 \rangle}$} & \multirow{3}{*}{Yes} & Presents an active steady state; & \multirow{3}{*}{\cite{Newman010:book}, \cite{Li20141042}, \cite{Chen2014196}}\\
&  			& $Y \xrightarrow{\delta} Z$    & & & Short-term immunity. & \\
&  			& $Z \xrightarrow{\gamma} X$    & & & & \\
\hline
\hline
\multirow{23}{*}{\vspace{-1cm}\hspace{0.2cm}\begin{rotate}{90}
\hbox{Rumor}
\end{rotate}\hspace{0.2cm}} 
& \multirow{3}{*}{Maki -- Thompson} & $Y+X \xrightarrow{\lambda} Y+Y$ & \multirow{3}{*}{$\frac{\lambda}{\alpha} > 0$} & \multirow{3}{*}{Yes} & Absorbing state, $R$; & \multirow{3}{*}{\cite{makithompson1973}, \cite{Barrat08:book}, \cite{Lebensztayn2011}}\\
&   		      & $Y+Y \xrightarrow{\alpha} Z+Y$  & &  			 & Directed contact. & \\
&   		      & $Y+Z \xrightarrow{\alpha} Z+Z$  & &  			 & & \\
\cline{2-7}
& \multirow{4}{*}{Nekovee et al.} & $Y+X \xrightarrow{\lambda} Y+Y$ & \multirow{4}{*}{$\frac{\lambda}{\delta} \geq \frac{\langle k \rangle}{\langle k^2 \rangle}$} & \multirow{4}{*}{Yes} & Absorbing state, $R$; & \multirow{4}{*}{\cite{Nekovee2007}}\\
&   		      & $Y+Y \xrightarrow{\alpha} Z+Y$  & &  			 & Presents a  & \\
&   		      & $Y+Z \xrightarrow{\alpha} Z+Z$  & &  			 & ``lost of interest'' mechanism. & \\
&   		      & $S \xrightarrow{\delta} R$  & &  			 & & \\
\cline{2-7}
& \multirow{4}{*}{Borge et al.~\tnote{3}} \hspace{0.1cm} & $Y+X \xrightarrow{\lambda \eta} Y+Y$ & \multirow{4}{*}{--} & \multirow{4}{*}{Yes} & Absorbing state, $R$; & \multirow{4}{*}{\cite{Borge-Holthoefer2012}}\\
&   		      & $Y+X \xrightarrow{\lambda (1-\eta)} Z+Y$  & &  		 & Presents an & \\
&   		      & $Y+Y \xrightarrow{\alpha} Z+Z$  & &  			 & apathy mechanism; & \\
&   		      & $Y+Z \xrightarrow{\alpha} Z+Z$  & &  			 & Models activity. & \\
\cline{2-7}
& \multirow{9}{*}{Kawachi et al.~\tnote{4}} & $Y+X \xrightarrow{\alpha_{xy} \theta_{xy}} Y+Y$ & \multirow{9}{*}{--} & \multirow{9}{*}{No} &  & \multirow{9}{*}{\cite{Kawachi200855}}\\
& & $Y+X \xrightarrow{\alpha_{xy} (1 - \theta_{xy})} Y+Z$ & & & & \\

& & $Z+X \xrightarrow{\alpha_{xz} \theta_{xz}} Z+Y$ & & & & \\
& & $Z+X \xrightarrow{\alpha_{xz} (1 - \theta_{xz})} Z+Z$ & & & Presents an active steady state. & \\

& & $Y+Y \xrightarrow{\beta} Y+Z$ & & &  & \\

& & $Z+Y \xrightarrow{\gamma} Z+Z$ & & & & \\

& & $Z+Z \xrightarrow{\lambda p} Z+X$& & & & \\

& & $Y \xrightarrow{\eta_y} X$& & & & \\
& & $Z \xrightarrow{\eta_z} X$& & & & \\
\hline
\end{tabular}
\label{tab:literature}
\begin{tablenotes}
\item[1]{This expression is obtained in details in~\cite{Newman010:book}. In~\cite{Barrat08:book}, the authors show another compartmental based approach, which yields $\frac{\lambda}{\delta} > \frac{\langle k \rangle}{\langle k^2 \rangle - \langle k \rangle}$. The first follows a quenched mean field (QMF) approach, where the process takes place on a fixed network, while the second expression is obtained considering the degree-based mean field (DMF) approach, where we assume that every node with the same degree is statistically equivalent.}
\item[2]{This expression is obtained in details in~\cite{Newman010:book}. In~\cite{Barrat08:book} the authors show that $\frac{\lambda}{\delta} > \frac{\langle k \rangle}{\langle k^2 \rangle}$, that is completely analogous to the expression in the table, since $\Lambda_{max} = \frac{\langle k^2 \rangle}{\langle k \rangle}$ for a random network generated by the configuration model~\cite{ChungPNAS2003} considering scale-free networks in which $P(k) \sim k^{-\zeta}$ and $2 < \zeta < \frac{2}{5}$.}
\item[3]{Considering the Model II in~\cite{Borge-Holthoefer2012}, which takes into account the apathy of the individuals.}
\item[4]{The authors considered even more interactions, but did not evaluate most of the possibilities numerically. In addition, the transitions follow the notation used in~\cite{Kawachi200855} and are rates, not necessarily probabilities, since the authors follow a continuous time approach.}
\end{tablenotes}
\end{threeparttable}
\end{table*}


\section{Previous spreading models} \label{sec:review}

Spreading processes in networks, such as the propagation of diseases or rumors, are based on (i) spontaneous processes, in which each node changes its state without any external interference, or on (ii) contact based processes, in which the state of each individual changes due to contact with its neighbors. In the simplest case, each element performs only one contact per time unit. This process is called contact process (CP). On the other hand, when every neighbor of a node is contacted in one time step, the process is called reactive process (RP). Intermediate situations between CP and RP can be defined by considering a parameter that defines the activity of each node~\cite{Gomez2010, Gomez2011}. Here we only consider the CP and RP schemes. 

We assume that a contagion process could refer to a disease or a rumor that spreads on top of complex networked systems. Although one can study other variants of disease compartmental models, here we explore epidemic models in which there are at most three different dynamical states: (i) susceptible, which accounts for subjects that do not have the disease; (ii) infected, which denotes individuals who have the disease and are transmitting it, and (iii) recovered or removed, which stands for immunized or cured subjects. The simplest epidemic model is the susceptible-infected (SI), which models a disease with no cure. On the other hand, in the susceptible-infected-susceptible (SIS) model, individuals recover but do not acquire immunity in front of the disease, and, therefore, they can catch the disease again, that is, once recovered, they go back to the susceptible state. At variance with the SIS scenario, in the susceptible-infected-recovered (SIR) model, the recovery is permanent, i.e., each individual acquires lifetime immunity. Above the critical point, the SIS model exhibits a steady state, where the number of susceptible and infective individuals are constant proportions of the population. More precisely, above the threshold, the SIS model presents a meta-state, where there is a probability large than zero that a certain fraction of the population is infected. It is worth mentioning that the only absorbing state of such a dynamics is the disease free configuration. In contrast, in the same regime, the SIR model presents an absorbing state when the fraction of infective is zero, being the number of recovered larger than zero. 

The rumor models here considered also have three classes: (i) ignorant, which represents individuals that have not heard the rumor; (ii) spreader, which refers to nodes that are aware of the rumor and are actively spreading it further, and (iii) stifler, who are those subjects that already know the rumor, but are not disseminating it any longer. The two main models of rumor spreading are due to Daley and Kendal (DK)~\cite{daley1964,daley1965,Barrat08:book} and Maki and Thompson (MT)~\cite{makithompson1973, Barrat08:book}. In the DK model, an edge of a network is sampled randomly, characterizing a contact. If this edge has a spreader and an ignorant, then the rumor is propagated from the spreader to the ignorant at a rate $\lambda$. On the other hand, if this edge is composed by two spreaders (or a spreader and a stifler), then the two spreaders turn into stiflers at a rate $\alpha$. In the MT model, on the other hand, directed contacts are accounted for. Thus, if a spreader contacts a stifler or another spreader, then only the initial spreader, who performed the contact, becomes a stifler at the same rate $\alpha$. The traditional DK and MT models have been adapted to allow for heterogeneous contact patterns~\cite{moreno2004} and spontaneous lost of interest~\cite{Nekovee2007, Borge-Holthoefer2012}.

For the sake of generality, given that we shall aim at developing a general model for spreading processes $-$no matter if a disease or a piece of information$-$, we consider throughout this paper the following states: (i) susceptible or ignorant, which we denote by state $X$, (ii) infected or spreader, which is represented by state $Y$; and (iii) recovered or stifler, corresponding to state $Z$. The reader is also referred to Table~\ref{tab:literature}, where we have summarized the main rumor and epidemic models studied so far. The transitions between states in each model and the thresholds, when available, are also shown for completeness.


\section{The general model} \label{sec:model}

\begin{figure*}[!t]
\begin{center}
\includegraphics[width=0.85\textwidth]{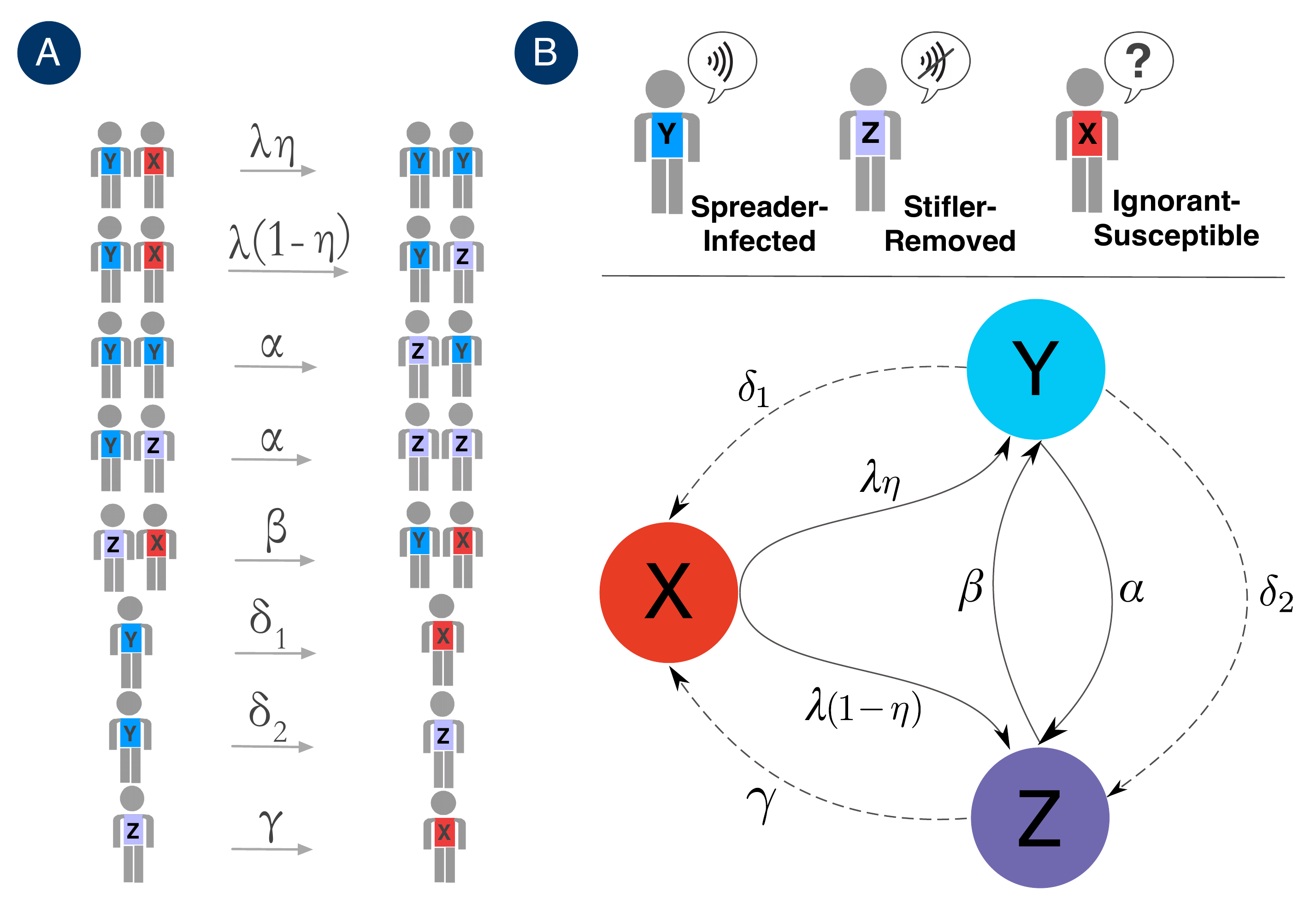}
\caption{A simplified diagram of all possible transitions between the three different dynamical states in our model. $X$, $Y$ and $Z$ stand for ignorant or susceptible, spreader and stifler or recovered, respectively. Most of the transitions involve interactions between two individuals, however, spontaneous ones are also allowed (represented by dashed lines).}
\label{Fig:Diagram}
\end{center}
\end{figure*}

Here we introduce a general model that captures all the features of the models sketched in the previous Section~\ref{sec:review}. We consider a spreading dynamics on a population of $N$ individuals whose contact pattern is given by a network. The structure of the interaction graph is encoded by a network with adjacency matrix $\mathbf{A}$, where $A_{ij} = 1$ if there is a connection between nodes $i$ and $j$, and $A_{ij} = 0$ otherwise. We consider undirected networks, i.e., $A_{ij}=A_{ji}$, for all $i,j$. The spreading process described by the model could refer to the transmission of a rumor, a disease or information that could be disseminated (also, a malware over a communication network, innovations, etc). In order to be more clear and precise, we will use henceforth a terminology that mostly refers to information transmission, but we stress that depending on the transitions that one allows to take place, the dynamics could represent other spreading processes. Therefore, let's assume that a "piece of information" is being transmitted. An individual holding this piece of information and willing to spread it is called a spreader, whose state is represented by $Y$. On the other hand, a stifler, whose state is denoted by $Z$, is a subject who knows the information and does not spread it. An individual who is not aware of the information is called ignorant, and its state is denoted as $X$. In this way, the notation $X$, $Y$ and $Z$, refers to subjects (or system's elements like devices) that have not participated in the spreading process, that are active, and that have taken part on the dynamics but are not active any longer, respectively. Hence, when we deal with an infectious disease, $X$, $Y$ , and $Z$ are, respectively, the susceptible, infective, and removed states. 

The general model here described includes the transitions represented in Fig.~\ref{Fig:Diagram}. Specifically, at each time step, the spreading dynamics proceeds as follows for any given node $i$:

\begin{itemize}
\item[(i)] An individual in state $Y$ changes to state $X$ with probability $\delta_1$. This, for instance, refers to the case in which in a rumor process, a spreader forgets the rumor (meaning the piece of news becomes old and therefore $Y$ is ignorant again). It also represents transitions of the type infected$\rightarrow$susceptible in a SIS model.
\item[(ii)] An individual in state $Y$ changes to state $Z$ with probability $\delta_2$. For rumor processes, it represents the case in which a spreader spontaneously (that is, not as a result of interactions with others $Y$ or $Z$ individuals) loses the interest in further propagating the rumor and becomes stifler. It also accounts for transitions of the type infected$\rightarrow$recovered in a SIR model.
\item[(iii)] If transitions (i) and (ii) do not happen, which occurs with probability $(1 - \delta_1 - \delta_2)$, then the spreader $i$ interacts with its neighbors. The outcome of such interactions are:
\begin{itemize}
\item[(iii.1)] If the individual contacted (the one at the other end of the edge) is in state $X$, then the latter turns into $Y$ with probability $\lambda\eta$. This transition is the classical susceptible$\rightarrow$infected one in disease models for $\eta=1$. Traditional rumor models also include the very same transition (ignorant$\rightarrow$spreader, also for $\eta=1$), but here, in order to be as general as possible, we also consider that an ignorant $X$ can directly go to the stifler class $Z$ with probability $(1-\eta)$, thus the probability that a transition $X\rightarrow Z$ occurs is $\lambda(1-\eta)$. This mimics information dissemination in systems like Twitter, in which reading the tweet does not imply that the user spreads it $-$as a matter of fact, the most common situation is that in which the user does not retweet the piece of news~\cite{Borge-Holthoefer2012}.
\item[(iii.2)] On the contrary, if the neighbor of the spreader $i$ is in state $Y$ or $Z$, then $i$ turns into $Z$ with probability $\alpha$. Note that this spreader$\rightarrow$stifler transition is only found in rumor models and it is not present in disease contagion. 
\end{itemize}
\item[(iv)] Finally, at variance with traditional rumor models~\cite{makithompson1973,daleykendall1964}, we also ascribe an active role to stiflers $Z$. We assume that individuals in state $Z$ can go back to state $X$ with probability $\gamma$. This represents scenarios in which stiflers might spontaneously "forget" the rumor, thus becoming ignorants again. We also note that this transition can be identified with a recovered$\rightarrow$susceptible one in SIRS disease models, in which it is assumed that after some time, individuals might lose their acquired immunity and become susceptible to catch the same disease again. 
\item[(v)] If (iv) does not happen, which occurs with probability $(1-\gamma)$, then the node $i$ in state $Z$ interacts with its neighbors. If the contact is with a subject in state $X$ (ignorant), the stifler (state $Z$) might recover the interest on the rumor propagation and with probability $\beta$ turns into $Y$. This transition mimics cases in which an individual who knows the rumor but is not transmitting it, learns that the rumor is still active and new, and therefore starts spreading it again. 
\end{itemize}

In summary, our model presents the following transitions:
\begin{equation*}
\begin{array}{rcl}
Y + X & \xrightarrow{\lambda \eta} & Y + Y,\\[.2cm]
 Y + X & \xrightarrow{\lambda (1-\eta)} &Y + Z,\\[.2cm]
 Y + Y &\xrightarrow{\alpha} &Z + Y,\\[.2cm]
 Y + Z &\xrightarrow{\alpha} &Z + Z,\\[.2cm]
 Z + X &\xrightarrow{\beta} &Y + X,\\[.2cm]
 Y &\xrightarrow{\delta_1} &X,\\[.2cm]
 Y &\xrightarrow{\delta_2} &Z,\\[.2cm]
 Z &\xrightarrow{\gamma} &X,
\end{array}
\end{equation*}
and our next goal will be to describe a process with such transitions through a suitable Markov chain approach. 

\subsection{The Markov-chain formulation}

In order to describe the evolution of this phenomenon in a given network made up by a set of nodes $[N]:=\{1,2,\ldots,N\}$, and adjacency matrix ${\bf A}$, we construct a discrete time Markov chain $(\xi_t)_{t\geq 0}$ with state space $\mathcal{S}=\{(1,0,0),(0,1,0),(0,0,1)\}^{[N]}$. More precisely, we define $\xi_t :=\{(X_i(t),Y_i(t),Z_i(t)): i\in [N]\}$, where $X_i(t), Y_i(t)$ and $Z_i(t)$ are Bernoulli random variables indicating whether the node $i\in[N]$ is an ignorant (susceptible), a spreader (infected), or a stifler (recovered) at time $t$, respectively. So, $(1,0,0),(0,1,0)$ and $(0,0,1)$ represent the states $X$, $Y$ and $Z$, respectively. Each point $\xi \in \mathcal{S}$ is called configuration. To construct the Markov chain, we consider random objects defined on the same suitable probability space $(\Omega, \mathcal{F}, \mathbb{P})$, where $\Omega$ is the sample space,  $\mathcal{F}$ is a $\sigma$-algebra of subsets of $\Omega$, i.e., the set of events, and $P$ is a probability measure function. For every $i,j\in [N]$ we take the following independent collections, each of independent and identically distributed (i.i.d.) random variables

\begin{equation}\label{eq:collections}
\begin{array}{ccccc}
\text{Collection} &\hspace{.5cm} & \text{Associated to}&\hspace{.5cm} & \text{Influence a choice on}\\[.2cm]
\left\{U_i(t)\right\}_{t\geq 0} &\hspace{.5cm} & Y &\hspace{.5cm} & \text{ state }Y\\[.2cm]
\left\{I_i^{\gamma}(t)\right\}_{t\geq 0} &\hspace{.5cm} & Z&\hspace{.5cm} & \text{ state }Z\\[.2cm]
\left\{I_{ij}^{\lambda}(t)\right\}_{t\geq 0} & \hspace{.5cm}& Y+X&\hspace{.5cm} & \text{ state }Y\\[.2cm]
 \left\{I_i^{\alpha}(t)\right\}_{t\geq 0} &\hspace{.5cm} & Y+Y \text{ or }Y+Z &\hspace{.5cm} & \text{ state }Y\\[.2cm]
  \left\{I_i^{\beta}(t)\right\}_{t\geq 0} &\hspace{.5cm} & Z+X&\hspace{.5cm} &\text{ state }Z\\[.2cm]
  \left\{I_i^{\eta}(t)\right\}_{t\geq 0}&\hspace{.5cm}  &Y+X&\hspace{.5cm} &\text{ state }X
  \end{array}
\end{equation}

\noindent
where  $U_i(1)$ is a random variable uniformly distributed on $(0,1)$, and $I_i^{\gamma}(1), I_{ij}^{\lambda}(1), I_i^{\alpha}(1), I_i^{\beta}(1)$ and $I_i^{\eta}(1)$ are Bernoulli random variables with parameter $\gamma, \lambda, \alpha, \beta$ and $\eta$, respectively. In addition, for each node $i$, we consider a sequence of i.i.d. random objets uniformly distributed on the neighbors of $i$, i.e., a sequence $U_i^{nb}(1),U_i^{nb} (2),\ldots$ such that $P(U_i^{nb}(1)=j)=1/k_i$, for all $j\in [N]$ satisfying $A_{ij}=1$, where $k_i =\sum_{j\in [N]}A_{ij}$ is the number of neighbors of node $i$.  

The main idea is to define a stochastic process that evolves according to the realization of the random variables defined above. For example, let's think of a rumor process. If at a fixed time $t$ the node $i$ is in state $Y$, then it forgets the rumor if $Y_i(t)<\delta_1$ (transition (i) above), it loses the interest on the propagation whenever $\delta_1\leq Y_i(t)<\delta_1 + \delta_2$ (transition (ii) above), or contacts its neighbors when $Y_i(t)\geq \delta_1+\delta_2$ (i.e., transition (iii) above). In the last situation, if node $i$ contacts node $j$, and $j$ is in  state $X$, then the rumor is propagated from $i$ to $j$ if $I_{ij}^{\lambda}(t)=1$. On the other hand, if the contacted node $j$ is a spreader, then $i$ turns into a stifler, i.e., $I_{i}^{\alpha}(t)=1$. We can proceed in a similar way to represent the remaining transitions and interactions of the process. Given the above description, it is not difficult to see that the transitions of this stochastic process can be written, for each $i\in [N]$ and $t\geq 0$, as:

\begin{widetext}
\begin{equation} \label{eq:MC_transitions}
\begin{cases}
\begin{array}{rcl}
X_i(t+1) & = & X_i(t)  A_i(t) + Y_i(t)  {1}_{\{U_i^t<\delta_1\}} + Z_i(t)  I_{i}^{\gamma}(t) \\[.3cm]

Y_i(t+1) & = & X_i(t)  (1 - A_i(t)) I_i^{\eta}(t) +\\[.2cm]
	    && Y_i(t)  1_{\{U_i^t\geq \delta_1 + \delta_2\}} B_i(t) +\\[.2cm]
	    && Z_i(t)  (1 - I_i^{\gamma}(t)) (1 - C_i(t)),\\[.3cm]
	    
Z_i(t+1) &= & X_i(t)   (1 - A_i(t)) (1 - I_i^{\eta}(t)) + \\[.2cm]
		& & Y_i(t) \left(1_{\{U_i^t\geq \delta_1 + \delta_2\}} (1 -B_i(t)) + 1_{\{\delta_1 \leq U_i^t<  \delta_1 + \delta_2\}} \right)+\\[.2cm]
		 &&Z_i(t) (1 - I_i^{\gamma}(t))  C_i(t) , 
\end{array}
\end{cases}
\end{equation}
\end{widetext}
where $A_i(t)$, $B_i(t)$ and $C_i(t)$ are Bernoulli random variables indicating that a node $i$, given the influence of its neighbors, will not be informed, will not become a stifler, or will not recover the interest in rumor propagation, from time $t$ to time $t+1$, respectively. Observe that these variables depend on the contacts between node $i$ and its neighbors. 

Here we study two limiting cases, namely, the contact process (CP), in which each node performs only one contact per unit time; and the fully reactive process (RP), in which each vertex contacts all its neighbors at each time step. The contact based variables for the CP are given by 
\begin{equation} \label{eq:contact_cv}
\begin{cases}
A_i(t) = \prod_{j=1}^N \left[1-I_{ji}^{\lambda}(t) 1_{\{U_j^{nb}(t)=i\}} 1_{\{U_j^t \geq \delta_1 + \delta_2\}} Y_j(t)\right],\\[.2cm]
B_i(t) = 1 - \sum_{j=1}^N \left[I_i^{\alpha}(t) 1_{\{U_i^{nb} (t)=j\}} \left( Y_j(t) + Z_j(t) \right)\right],\\[.2cm]
C_i(t) = 1 - \sum_{j=1}^N \left[I_i^{\beta}(t)  1_{\{U_i^{nb} (t)=j\}} X_j(t)) \right].
\end{cases}
\end{equation}
On the other hand, the contact based variables for the RP are
\begin{equation} \label{eq:contact_rv}
\begin{cases}
A_i(t) = \prod_{j=1}^N \left[1-I_i^{\lambda}(t) A_{ji} 1_{\{U_j^t \geq \delta_1 + \delta_2\}} Y_j(t)\right],\\[.2cm]
B_i(t) = \prod_{j=1}^N \left[1-I_i^{\alpha}(t) A_{ij} \left( Y_j(t) + Z_j(t) \right)\right],\\[.2cm]
C_i(t) = \prod_{j=1}^N \left[1-I_i^{\beta}(t) A_{ij} X_j(t)) \right].
\end{cases}
\end{equation}
Therefore, the contact based variables of the node $i$, i.e., $A_i(t)$, $B_i(t)$ and $C_i(t)$ in equations~\eqref{eq:MC_transitions}, \eqref{eq:contact_cv} and~\eqref{eq:contact_rv} are dependent on the parameters $\lambda$, $\alpha$ and $\beta$, the state of its neighbors and the process, i.e., whether it is a CP or a RP. The variables $B_i(t)$ and $C_i(t)$ describe the feedback from the contacts to the node. Note that in rumor dynamics, interactions can also change the state of the spreading node, returning an immediate feedback, in contrast to epidemic spreading models~\cite{Satorras015} where the state of the spreader changes at a rate that is independent of the interaction network. These relations modeled by $B_i(t)$ and $C_i(t)$ are absent in the recent work by Stanoev et al.~\cite{Stanoev2014}, since their formalism does not allow for instantaneous feedback over node $i$. We also highlight that the CP and RP mimic different situations in, for instance, social networks. The CP models one-to-one communication, when the rumors spread in friendship networks, email networks or networks in which each individual interacts with only one contact at each time step. On the other hand, the RP is best suited for one-to-many information dissemination, as in Twitter-like networks, since when a user posts a tweet, all its followers receive this information. 

Finally, we also point out that Eq. \eqref{eq:MC_transitions} ensures the existence of a function $f$ such that 
\begin{equation*}\label{eq:MCrec}
\xi_{t+1}=f\left(\xi_t, \mathcal{I}_t\right),
\end{equation*}
where 
$$\mathcal{I}_t:=\bigcup_{i=1}^{N}\left\{U_i(t), I_i^{\gamma}(t), I_{ij}^{\lambda}(t), I_i^{\alpha}(t), I_i^{\beta}(t), I_i^{\eta}(t), U_i^{nb}(t) \right\}$$
is the collection of all the (independent) random variables at time $t$. This in turns implies that $(\xi_t)_{t\geq 0}$ is a discrete-time Markov chain.

\subsection{The mean-field approximation}

Although the above Markovian description of the model is exact, its analysis is rather complex. This is because, while the Markov chain $(\xi_t)_{t\geq 0}$ is defined from the realizations of independent random variables through \eqref{eq:MC_transitions}, the state of each node as well as of its neighbors have dependencies, whose complexity might change according to the network. We solve this difficulty by considering a mean-field approximation for the Markovian description. We denote the probabilities that a node $i$ is in state $X$, $Y$ and $Z$ at time $t$ by $p_i^X(t)$, $p_i^Y(t)$ and $p_i^Z(t)$, respectively, and note that
\begin{equation}
\begin{array}{rcccl}
p_i^X(t) &:= & \mathbb{P}\left(X_i (t)=1\right) & = & \mathbb{E}(X_i (t)),\\[.2cm]
 p_i^Y(t)&:= & \mathbb{P}\left(Y_i (t)=1\right) & = & \mathbb{E}(Y_i (t)),\\[.2cm]
  p_i^Z(t)& :=& \mathbb{P}\left(Z_i (t)=1\right) & = & \mathbb{E}(Z_i (t)),
\end{array}
\end{equation}
where $\mathbb{E}$ denotes expected values. Therefore, the mean-field approximation is obtained by considering expected values in equations~\eqref{eq:MC_transitions}, \eqref{eq:contact_cv} and~\eqref{eq:contact_rv}. We also assume that there are no dynamical correlations at first order. In other words, we assume that the expected values of variable pairs factorize. The resulting system of equations obtained for a node $i\in[N]$, for all $t\geq 0$, is given by
\begin{equation} \label{eq:rumor_evo}
\begin{cases}
p_i^X(t+1)  = & p_i^X(t)  a_i(t) + p_i^Y(t)  \delta_1 + p_i^Z(t)  \gamma \\[.2cm]
p_i^Y(t+1)  = & p_i^X(t)  (1 - a_i(t)) \eta +  \\[.2cm]
	    & p_i^Y(t)   (1 - \delta_1 - \delta_2) b_i(t) + \\[.2cm] 
	    & p_i^Z(t)  (1 - \gamma) (1 - c_i(t)),\\[.2cm]
p_i^Z(t+1)  = & p_i^X(t)   (1 - a_i(t)) (1 - \eta) + \\[.2cm]
		 & p_i^Y(t) (1 - \delta_1 - \delta_2) (1 -b_i(t))  + p_i^Y(t)\delta_2\\[.2cm]
		 & p_i^Z(t) (1 - \gamma)  c_i(t) , 
\end{cases}
\end{equation}
where $a_i(t)$, $b_i(t)$ and $c_i(t)$ are the probabilities that a node $i$, given a contact with its neighbors, remains at state $X$ (e.g., it's not informed), does not change to state $Z$ from state $Y$ (that is, will not become a stifler), or remains at state $Z$ (that is, will not recover the interest in rumor propagation) in the time interval from $t$ to $t+1$, respectively (see also Fig.\ref{Fig:Diagram}). These probabilities depend on the number of contacts per unit time. Thus, the contact based probabilities for the CP are given by 
\begin{equation} \label{eq:contact_cp}
\begin{cases}
a_i(t) = \prod_{j=1}^N \left[1-\lambda P_{ij} (1 - \delta_1 - \delta_2) p_j^Y(t)\right], \\[.2cm]
b_i(t) = 1 - \sum_{j=1}^N \left[\alpha P_{ij} \left( p_j^Y(t) + p_j^Z(t) \right)\right],  \\[.2cm]
c_i(t) = 1 - \sum_{j=1}^N \left[\beta  P_{ij} p_j^X(t)) \right],
\end{cases}
\end{equation}
where $P_{ij} := A_{ij}/k_i$.  On the other hand, the contact based probabilities for the RP are
\begin{equation} \label{eq:contact_rp}
\begin{cases}
a_i(t) = \prod_{j=1}^N \left[1-\lambda A_{ji} (1 - \delta_1 - \delta_2) p_j^Y(t)\right],  \\[.2cm]
b_i(t) = \prod_{j=1}^N \left[1-\alpha A_{ij} \left( p_j^Y(t) + p_j^Z(t) \right)\right], \\[.2cm]
c_i(t) = \prod_{j=1}^N \left[1-\beta A_{ij} p_j^X(t)) \right].
\end{cases}
\end{equation}

The system of equations~\eqref{eq:rumor_evo} describes the micro-state evolution of the system, i.e., the evolution of the probabilities of each node to be in a given state. The macro-state variables can be defined as the average of the individual probabilities, namely,
\begin{equation}
 \rho^{\ell} = \frac{1}{N} \sum_{i=1}^N p_i^{\ell}(\infty)
\end{equation}
where $p_i^{\ell}(\infty)$ is the probability of node $i$ to be in state $\ell$, with $\ell \in \{ X, Y, Z\}$, as $t\to \infty$. Note that the generalization for weighted and directed networks is straightforward. In weighted networks, it is necessary to consider the weight matrix whose rows must sum up to one. In directed networks, the degree must be substituted by the out-degree, $k_{out}$. For the CP case, the directed network must be an ergodic Markov chain. Another possibility of generalization is the consideration of heterogeneous parameters, which is treated naturally in the set of equations~\eqref{eq:rumor_evo},~\eqref{eq:contact_cp} and~\eqref{eq:contact_rp}. In the following analysis we assume that the parameters have the same values for every nodes. In addition, we consider all networks as undirected and unweighted.
 
\subsection{Homogeneously mixed population} \label{sec:model_complete}

\begin{figure}[t]
\begin{center}
\includegraphics[width=0.95\columnwidth]{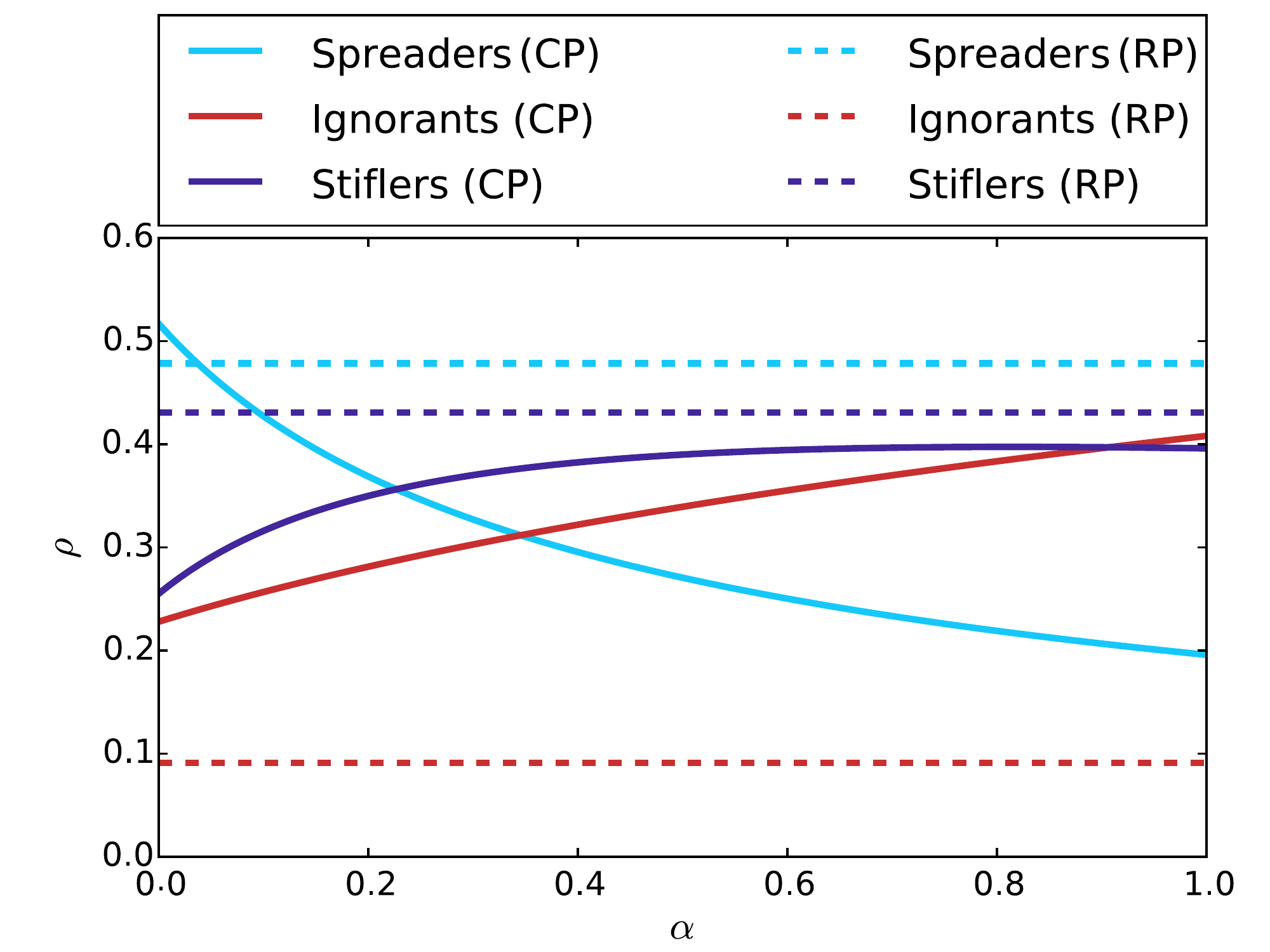}
\caption{Phase diagram of the steady state ($t\rightarrow \infty$) considering the complete infinity graph. The dynamical parameters have been set to $\delta_1 = \delta_2 = \gamma = 0.1$, $\lambda = 1$, $\beta = 0.5$ and $\eta = 1$.}
\label{Fig:MK_CP_Complete}
\end{center}
\end{figure}

For the sake of comparison and completeness, we next study our model on a homogeneously mixed population, i.e., we evaluate our model on a complete graph when the number of nodes goes to infinite. For a complete graph with $N$ nodes, and for all $i\in [N]$, we have that $A_{ij}=1$, for all $j\in[N]\setminus \{i\}$, which implies $P_{ij} = \frac{1}{N-1}$. We obtain for the CP, from the set of equations~\eqref{eq:contact_cp}, and by taking the thermodynamic limit $N \rightarrow \infty$, that the probabilities that a node does not perform a transition to another state after a pairwise interaction at time $t$ are given by
\begin{equation} \label{Eq:infinityCP}
\begin{cases}
a(t) = e^{ \left[-\lambda (1 - \delta_1 - \delta_2) p^Y(t)\right]}, \\
b(t) = 1 - \alpha \left[p^Y(t) + p^Z(t) \right], \\
c(t) = 1 - \beta  p^X(t).
\end{cases}
\end{equation}
Note that we omit the index $i$ in our notation, since the probabilities are the same for every node. We observe that the probability $a(t)$ decays exponentially fast as a function of $p^Y(t)$, with a parameter $\tau^{-1} = \lambda (1 - \delta_1 - \delta_2)$. After including the probabilities~\eqref{Eq:infinityCP} into equation~\eqref{eq:rumor_evo}, we obtain the steady state behavior of this simplified model shown in Figure~\ref{Fig:MK_CP_Complete}. As it can be seen, the expected values for the densities of ignorant, spreaders and stiflers depend on $\alpha$, which is the tuning parameter in this scenario. 

Interesting enough, the behavior of the RP scenario is radically different. For the RP, after taking the limit $N \rightarrow \infty$ in Eq. \eqref{eq:contact_rp}, we get that $a(t) = 0$, $b(t) = 0$ and $c(t) = 0$, since those probabilities are based on the product of infinite terms with absolute values less than unity. As a consequence, the system of equations~\eqref{eq:rumor_evo} for the RP becomes in the limit $N \rightarrow \infty$
\begin{equation} \label{eq:rumor_rp_complete}
\begin{cases}
p^X(t+1)  = &  p^Y(t)  \delta_1 + p^Z(t)  \gamma, \\
p^Y(t+1)  = & p^X(t)  \eta  +  p^Z(t) (1 - \gamma), \\
p^Z(t+1)  = & p^X(t)  (1 - \eta)  +  p^Y(t)  (1 - \delta_1).
\end{cases}
\end{equation}
The evolution of the fraction of spreaders, ignorants and stiflers are also shown in Figure~\ref{Fig:MK_CP_Complete}. Note that these quantities are constant in time in sharp contrast to the results for the CP. 

\section{Analytical Analyses} \label{sec:Analytical}

In this section, we evaluate our model in terms of its early time behavior and perform the steady state analysis.  We derivate some closed expressions for the thresholds of the CP and RP cases in terms of the spectra of the probability transition and the adjacency matrices, respectively. 

\subsection{Early time behavior} \label{sec:Early}

Consider the first steps of the process. For a small time $\tilde{t}$ we may assume that $p_{i}^{Y}(t)\approx \epsilon_{i}^Y$, and $p_{i}^{Z}(t)\approx \epsilon_{i}^Z$, for any $t\leq \tilde{t}$, where $\epsilon_{i}^Y$ and $\epsilon_{i}^Z$ are constants such that $0 \leq \epsilon_{i}^Y \ll 1$, and $0 \leq \epsilon_{i}^Z \ll 1$. This in turns implies that $p_{i}^{X}(t)=1-p_{i}^{Y}(t)-p_{i}^{Z}(t)\approx 1-\epsilon_{i}^Y - \epsilon_{i}^Z \approx 1$. This approximation takes into account that the information spreading starts from at most a few spreaders.
Neglecting second-order terms for the RP in equation~\eqref{eq:contact_rp} we obtain, for $t\leq \tilde{t}$
\begin{equation} \label{eq:approx_rp}
\begin{cases}
a_i(t) \approx 1 - \sum_{j=1}^N \left[\lambda {A_{ji}} (1 - \delta_1 - \delta_2) p^{Y}_j(t)\right], \\
b_i(t) \approx 1 - \sum_{j=1}^N \left[\alpha {A_{ij}} \left( p^{Y}_j(t) + p^{Z}_j(t) \right)\right], \\
c_i(t) \approx 1 - \sum_{j=1}^N \left(\beta {A_{ij}} p^{X}_j(t)) \right).
\end{cases}
\end{equation}
Notice that 
\begin{equation} \label{eq:approx_u_rp}
c_i(t) \approx 1 - \beta k_i,
\end{equation}
since $p_i^X(t) \approx 1$, for $i = 1, 2, \ldots, N$. Substituting equations~\eqref{eq:approx_rp} and~\eqref{eq:approx_u_rp} in the system of equations~\eqref{eq:rumor_evo}, we obtain, for the equations of $p^{X}_i$ and $p^{Y}_i$, 
\begin{widetext}
\begin{equation} \label{eq:eps_rp}
\begin{cases}
 -\lambda (1 - \delta_1 - \delta_2) \sum_{j=1}^N A_{ij} \epsilon_{j}^Y + \delta_1 \epsilon_{i}^Y + \gamma \epsilon_{i}^Z \approx 0, \\
 \eta \lambda (1 - \delta_1 - \delta_2) \sum_{j=1}^N A_{ij} \epsilon_{j}^Y +  (1 - \delta_1 - \delta_2) \epsilon_{i}^Y + \beta k_i (1 - \gamma) \epsilon_{i}^Z \approx \epsilon_{i}^Y,
\end{cases}
\end{equation}
\end{widetext}
where $\epsilon_j^Y + \epsilon_j^Z \approx 0$, $\forall j = 1,2, \ldots, N$. Isolating $\epsilon_{i}^Z$ on the second equation, we get
\begin{equation} \label{eq:epsR}
 \epsilon_{i}^Z = \frac{ (\delta_1 + \delta_2) \epsilon_{i}^Y - \eta \lambda (1 - \delta_1 - \delta_2) \sum_{j=1}^N A_{ij} \epsilon_{j}^Y }{\beta k_i (1 - \gamma)} ,
\end{equation}
which describes $\epsilon_{i}^Z$ as a combination of $\epsilon^Y_{j}$, $\forall j = 1, 2, ... N$. Substituting~equation~\eqref{eq:epsR} on the first equation of~\eqref{eq:eps_rp} we have
\begin{widetext}
\begin{equation}
 \left( -1 - \frac{\eta \gamma}{\beta k_i (1 - \gamma)} \right) \lambda (1 - \delta_1 - \delta_2) \sum_{j=1}^N A_{ij} \epsilon_{j}^Y + \delta_1 \epsilon_{i}^Y + \gamma \frac{\epsilon_{i}^Y 
\left(\delta_1 + \delta_2 \right)}{\beta k_i (1 - \gamma)} \approx 0,
\end{equation}
\end{widetext}
whose factorization is 
\begin{equation} \label{eq:final_early}
 \sum_{j=1}^N \left[ A_{ij} - \delta_{ij} \left( \frac{\gamma \left(\delta_1 + \delta_2 \right) + \delta_1 (1 - \gamma) \beta k_i }{\lambda (1 - \delta_1 - \delta_2) \left((1 - \gamma)\beta k_i 
+ \gamma \eta \right)} \right) \right] \epsilon_{i}^Y  \approx 0.
\end{equation}
Notice that equation~\eqref{eq:final_early} does not decouple the structure and the dynamics of the system, since it is not possible to isolate the terms depending on $k_i$. However, it is possible to evaluate the threshold for $\beta(1 - \gamma) \approx 0$. In this case, the system has nontrivial solutions when $\left( \frac{\delta_1 + \delta_2}{\eta \lambda (1 - \delta_1 - \delta_2)} \right)$ is an eigenvalue of $A$. Thus, the critical value, which depends on the largest eigenvalue of the adjacency matrix $\mathbf{A}$, is given by
\begin{equation} \label{eq:thr_rp}
 \Lambda_{max}^A \approx \left( \frac{\delta_1 + \delta_2}{\eta \lambda (1 - \delta_1 - \delta_2)} \right)_c,
\end{equation}
where $(\cdotp)_c$ denotes the critical point. Considering the critical point as a function of $\lambda$, i.e., assuming that the other dynamical parameters are fixed, the threshold vanishes in the thermodynamic limit for scale-free networks with a divergent second moment, similarly to epidemic spreading~\cite{Pastor-Satorras2001}. 

\begin{figure}[t]
\begin{center}
\includegraphics[width=0.98\columnwidth]{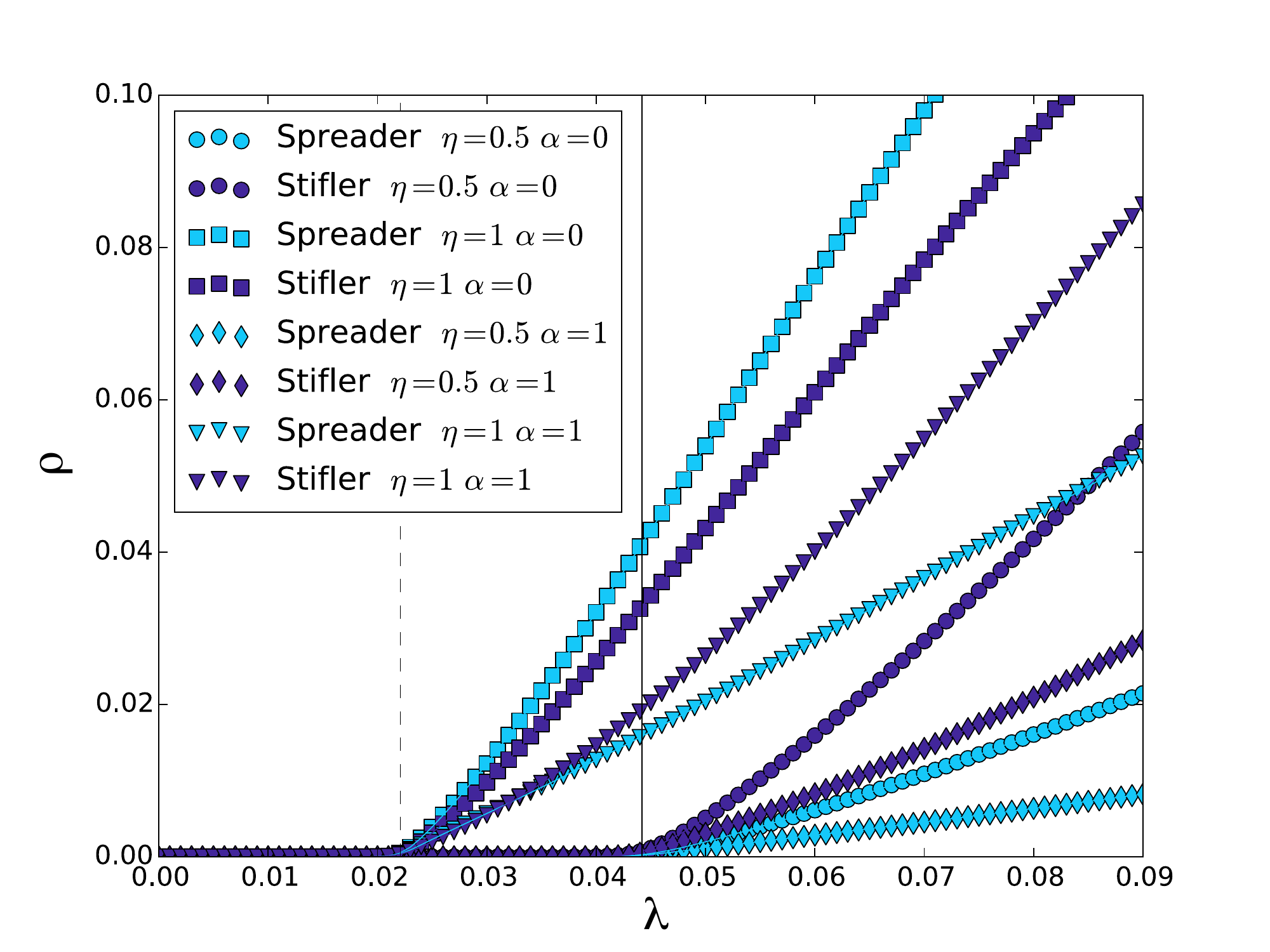}
\caption{Numerical evaluation of equation~\eqref{eq:rumor_evo} considering the RP for $\beta = 0$, $\delta_1 = \gamma = 0.25$, $\delta_2 = 0.2$ and varying the spreading probability $\lambda$. The network considered is a scale-free network with $\zeta \approx 2.7$, $N = 10^4$ and $\langle k \rangle \approx 10$. The dashed line indicates the critical point (Eq.~(\eqref{eq:thr_rp})) for $\eta = 1$, whereas the continuous line corresponds to the critical value for the case in which $\eta = 0.5$.}
\label{Fig:T_RP}
\end{center}
\end{figure}

The numerical evaluation of equation~\eqref{eq:thr_rp} is shown in Fig~\ref{Fig:T_RP}. As it can be seen, there is a good agreement between the theoretical predictions (equation~\eqref{eq:thr_rp}) and the numerical solution of the system~\eqref{eq:rumor_evo}. The calculation was performed for a scale-free network with $\zeta \approx 2.7$, $N = 10^4$ and $\langle k\rangle \approx 10$. As initial conditions, we considered that each node is set as a spreader with uniform probability $Y_i(0) = 0.01$, $\forall i \in \{ 1, 2, ... N\}$. The values of $\lambda$ in the figure are close to zero, since $\lambda$ depends on the inverse of the leading eigenvalue. We can also see that the limiting cases for $\alpha$ do not change the critical point. Furthermore, the density of nodes that holds the rumor, i.e., of spreaders and stiflers, is higher for $\alpha = 0$ and $\eta = 1$, while the lowest density is obtained for $\alpha = 1$ and $\eta = 0.5$. Such observation reinforces the role of stifling rates, similarly to what happens in classical rumor models.

\begin{figure}[t]
\begin{center}
\includegraphics[width=0.98\columnwidth]{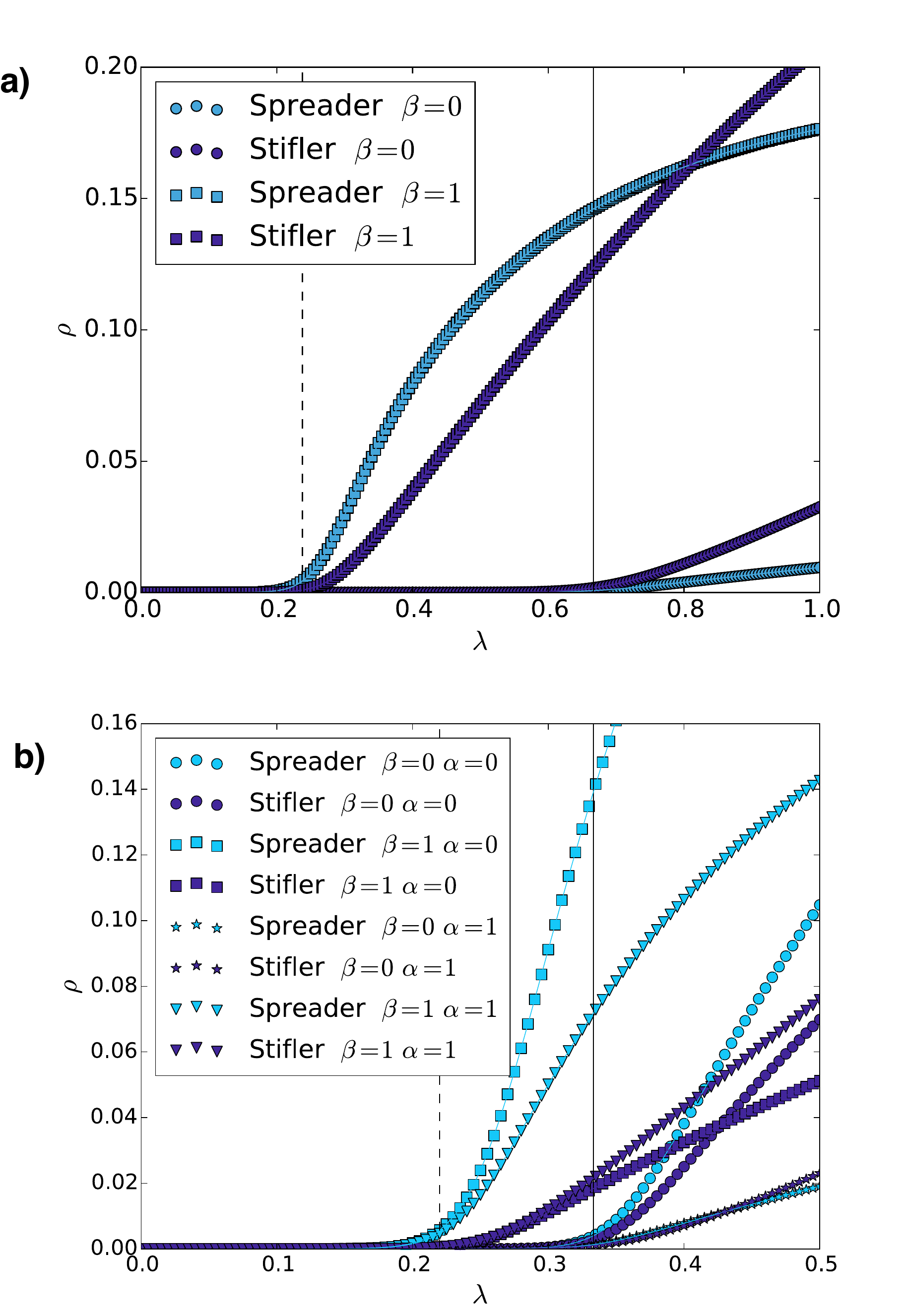}
\caption{Numerical evaluation of equation~\eqref{eq:rumor_evo} considering the CP with parameters $\delta_1  = \gamma = 0.15$, $\delta_2 = 0.1$ and varying the spreading probability $\lambda$. The network considered is a scale-free network with $\zeta \approx 2.7$, $N = 10^4$ and $\langle k \rangle \approx 100$. In panel (a) $\eta = 0.5$ and $\alpha = 1$, whereas in (b) we have set $\eta = 1$ and also explored the limiting cases of $\alpha=0$ and $\alpha=1$. The vertical lines correspond to the critical point Eq.(\ref{eq:thr_rp}) evaluated for the parameters used here.}
\label{Fig:T_CP}
\end{center}
\end{figure}

The approximation of the critical point in the RP is restricted by $\beta(1 - \gamma) \approx 0$, since there is an explicit dependence on the degree $k_i$. However, for the CP the critical point	can be evaluated without any constraint. The contact based probability in equation \eqref{eq:contact_cp} can be obtained using the same set of equations as in the RP (equation~\eqref{eq:approx_rp}). Notice that only the first equation, i.e., that for $a_i(t)$, is an approximation, whereas the equations for $b_i(t)$ and $c_i(t)$ are exact. Following the same approach as for the RP, the expression for the critical point for the CP is similar to that in equation~\eqref{eq:final_early} --- the only two changes are (i) the use of the matrix $\mathbf{P}$ instead of $\mathbf{A}$ and (ii) $\sum_j P_{ij} = 1, \forall j = 1,2, ... N$. Another important result is that the leading eigenvalue of the transition probability matrix is always equal to unity~\cite{Mieghem:2011}. In this way, the critical point for the CP is given by
\begin{equation} \label{eq:thr_cp}
 \Lambda_{max}^P = 1 \approx \left( \frac{\gamma \left(\delta_1 + \delta_2 \right) + \delta_1 (1 - \gamma) \beta }{\lambda (1 - \delta_1 - \delta_2) \left((1 - \gamma)\beta + \gamma \eta \right)} \right)_c.
\end{equation}
Note that such expression does not depend on the network structure, since the leading eigenvalue $\Lambda_{max}^P$ is the same for every connected network.  

The numerical evaluation of the expression~\eqref{eq:thr_cp} is shown in Fig.~\ref{Fig:T_CP} considering a scale-free network with $\zeta \approx 2.7$, $N = 10^4$ and $\langle k \rangle \approx 100$. In (a) we consider $\eta = 0.5$ and $\alpha = 1$, while in (b) we assume $\eta = 1$. Similarly to the RP case, here we also observe a very good agreement between the theoretical results and the numerical evaluation of the system~\eqref{eq:rumor_evo}  for all parameter values explored. The conclusions thus are similar as for the RP scenario, being the only difference the average degree of the network, a dependency that we shall analyze in more details later on. 

\begin{figure}[t]
\begin{center}
\includegraphics[width=0.98\columnwidth]{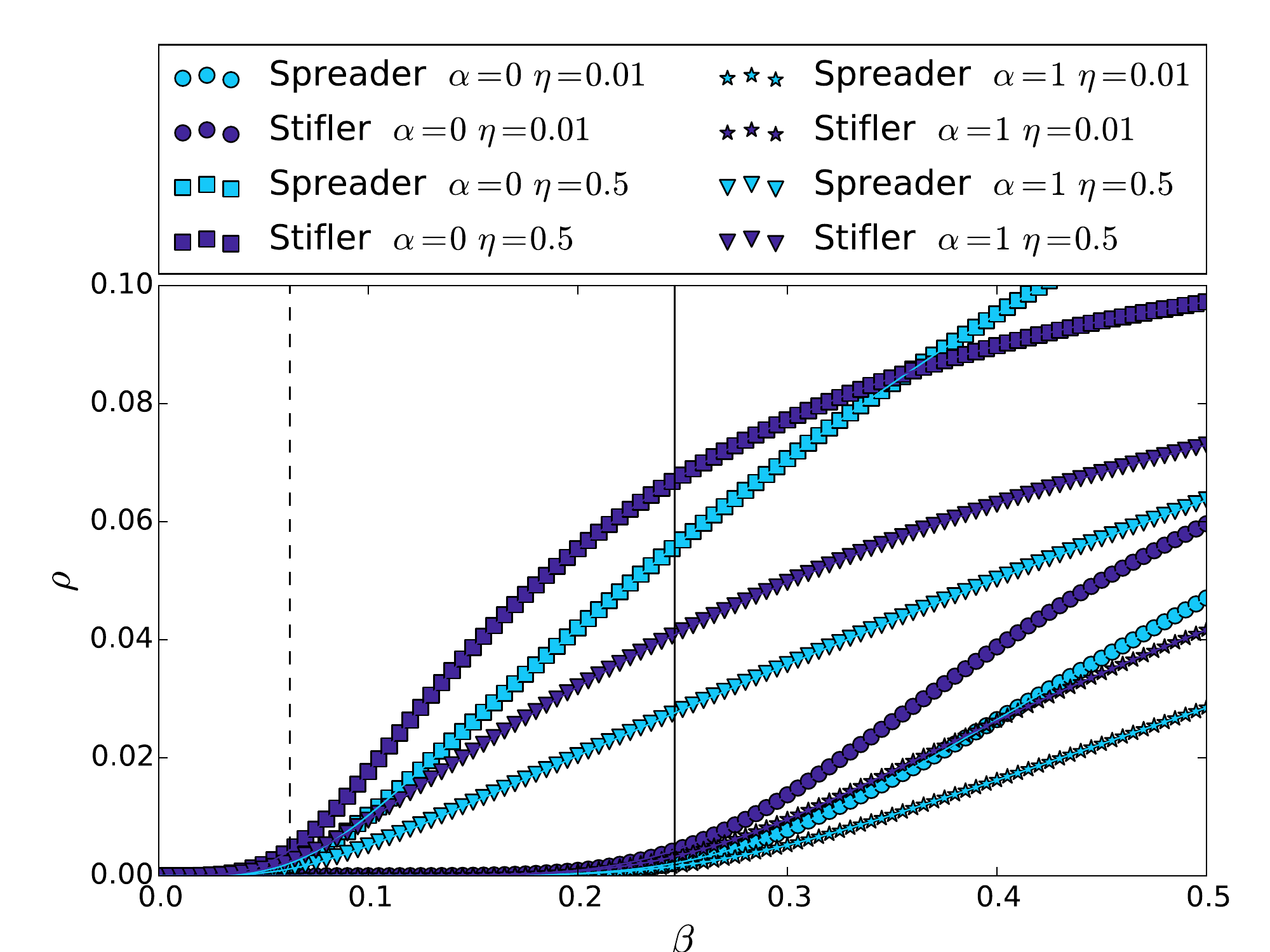}
\caption{Numerical evaluation of equation~\eqref{eq:rumor_evo} considering the CP and, without loss of generality, fixing the parameters $\delta_1 = \gamma = \delta_2 = 0.2$, $\lambda = 1$ and varying the parameter $\beta$. The network considered is a scale-free network with $\zeta \approx 2.7$, $N = 10^4$ and $\langle k \rangle \approx 100$.}
\label{Fig:T_CPbeta}
\end{center}
\end{figure}

The parameters of the model (with the exception of $\alpha$ that does not affect first order terms) can be used as control parameters of the system's dynamics. In particular, it is interesting to analyze the effect of $\beta$, which has been introduced here to account for the possibility of recovering the interest in the rumor. Isolating $\beta$ on equation~\eqref{eq:thr_cp} we obtain
\begin{equation} \label{eq:thr_cp_beta}
 \beta_c \approx \frac{\gamma (\delta_1 + \delta_2 - \eta \lambda (1 - \delta_1 - \delta_2))}{\lambda (1 - \delta_1 - \delta_2) - \delta_1 (1 - \gamma)},
\end{equation}
which defines the minimum value of $\beta$ that allows for the spreading of the rumor to a macroscopic fraction of the population for a given spreading rate $\lambda$ $-$as it is the case in most applications. Figure~\ref{Fig:T_CPbeta} shows the evaluation of equation~\eqref{eq:rumor_evo} for the CP near the critical point considering a scale-free network with $\zeta \approx 2.7$, $N = 10^4$ and $\langle k~\rangle~\approx~100$, and, without loss of generality, fixing the parameters $\delta_1 = \gamma = \delta_2 = 0.2$, $\lambda = 1$ and varying the parameter $\beta$. Similarly to the results shown in figures~\ref{Fig:T_RP} and~\ref{Fig:T_CP}, our approximations agree with the simulated values. 

As for the other important parameter, $\alpha$, that defines at which rate spreaders decay into stiflers after $Y-Y$ or $Y-Z$ interactions, we stress that any approximation neglecting second-order terms does not involve the parameter $\alpha$, since it controls the stifling rate on the expression for $b_i(t)$. In other words, $b_i$ is a conditional probability, and it is multiplied by the probability of an individual being an spreader. Thus, $\alpha$ does not affect the thresholds for the RP, since it always appears on second order terms (see Fig~\ref{Fig:T_RP}). The same occurs for the CP, as shown in Fig.~\ref{Fig:T_CP}. Note that while the macroscopic behavior of the system (the expected final densities of the different classes) is different for each parameter analyzed, the threshold is always the same. An equivalent analysis was done in~\cite{Nekovee2007}, where the authors proposed a model with a rate similar to $\delta_2$ (in our model) and showed the independence of the threshold on the parameter $\alpha$.

\subsection{Steady state analysis} \label{sec:Steady}

On the previous section we analyzed the early time evolution of the system aiming at finding its critical point. In this section, for the sake of completeness, we perform a similar analysis for the steady state solution. An analogous approach was used in~\cite{Gomez2010}, but for the specific case of a SIS model. At the steady state, we can assume, for $t$ large enough,  $p_i^{I}(t) \approx \pi_i^{I}$, $p_i^{Y}(t) \approx \pi_i^{Y}$ and $p_i^{R}(t) \approx \pi_i^{R}$, for $i = 1,2, \ldots, N$. Neglecting second-order terms, inserting~\eqref{eq:approx_rp} on the system~\eqref{eq:rumor_evo} and after some algebra, the steady-state solution is 
\begin{widetext}
\begin{equation} \label{eq:Steady}
\begin{cases}
\pi_i^{I} = \frac{(\delta_1 + \delta_2 \gamma + \delta_1 (-1 + \gamma) c_i)}{( \delta_2 (1 + \gamma - a_i) +   \delta_1 (2 + \eta (-1 + a_i) - a_i + (-1 + \gamma) c_i) - (-1 + a_i) (1 - c_i + \gamma (-1 + \eta + c_i)))} \\[.2cm]
\pi_i^{Y} = \frac{(-1 + a_i) (1 - c_i + \gamma (-1 + \eta + c_i))}{( \delta_2 (-1 - \gamma + a_i) + \delta_1 (-2 + \eta + a_i - \eta a_i + c_i - \gamma c_i) + (-1 + a_i) (1 - c_i + \gamma (-1 + \eta + c_i)))} \\[.2cm]
\pi_i^{R} = \frac{-(((\delta_1 + \delta_2 - \delta_1 \eta) (-1 + a_i))}{(  \delta_2 (1 + \gamma - a_i) +    \delta_1 (2 + \eta (-1 + a_i) - a_i + (-1 + \gamma) c_i) - (-1 + a_i) (1 - c_i + \gamma (-1 + \eta + c_i))))}.
\end{cases}
\end{equation}
\end{widetext}
where the time dependence does not appear any more.

Considering that $Y_i = \epsilon_i$, where $0 \leq \epsilon_i \ll 1$, and after substitution in the second equation of the system~\eqref{eq:Steady}, we obtain
\begin{widetext}
\begin{equation}
\epsilon_i ( \delta_2 \gamma + \delta_1 (c_i - \gamma c_i - 1)) = (a_i -1) (1 - c_i + \gamma (\eta + c_i - 1)),
\end{equation}
\end{widetext}
Taking into account the approximation for $a_i(t)$ in equation~\eqref{eq:approx_rp}, we have
\begin{multline}
\epsilon_i \left( \delta_2 \gamma + \delta_1 (k_i \beta (\gamma - 1) - \gamma) \right) =\\
 -\lambda (1 - \delta_1 - \delta_2) ((1 - \gamma) k_i \beta + \gamma \eta) \sum_{j=1}^N A_{ij} \epsilon_{j},
\end{multline}
which is the same equation as ~\eqref{eq:final_early}. This result is for the RP process. For the CP case, we only need to change the adjacency matrix $\mathbf{A}$ by the transition probability matrix $\mathbf{P}$. It is also worth highlighting that for an arbitrary choice of model parameters, in general, and at variance with classical rumor models, there is no absorbing state corresponding to the absence of spreaders. 

\section{Special cases} \label{sec:special}

The model proposed includes new transitions, formulates rumor dynamics in terms of discrete Markov chains and generalizes several previous spreading models (see Section~\ref{sec:review}). It is instructive to show how to obtain some of the main epidemic and rumor models from our approach, which we do next.

\subsection{Disease spreading}

We can obtain a particular case of the SIS model by setting $\eta = 1$, $\delta_2 = \gamma = 0$, $\beta = 0$ and $\alpha = 0$ in our model (see Figure~\ref{Fig:Diagram}). This model is not exactly a SIS model, since we assume that each node cannot spread the disease and become susceptible at the same time step. Such modification implies that the parameters used on the traditional models change. However, in both cases the SIS dynamics can be studied in terms of the final fraction of spreaders. From equation~\eqref{eq:thr_rp}, the epidemic threshold is given by
\begin{equation} 
 \Lambda_{max} \approx \left( \frac{\delta_1}{\lambda (1 - \delta_1)} \right)_c.
\end{equation}
Comparing with the results in Table~\ref{tab:literature}, the recovery probability in this SIS model is thus given by $\delta = \frac{\delta_1}{(1 - \delta_1)}$, while the spreading probability is $\lambda$.

In~\cite{Gomez2010}, the authors used a discrete time Markov chain approach to model the SIS dynamics considering a reinfection term and a parameter that allows to explore a family of contact-based scenarios, including as limiting cases the CP and the fully RP. Concerning the reinfection, our model does not include such a feature, however we note that reinfection within the same time step is rarely taken into account in SIS like models. On the other hand, the second ingredient can be easily incorporated into our framework by exchanging the matrix $\mathbf{A}$ by the matrix
\begin{equation}
 R_{ij} = 1 - \left(1 - \frac{A_{ij}}{k_i} \right)^{\psi_{i}}
\end{equation}
where $\psi_{i}$ is the activity parameter. Observe that the limiting cases are obtained setting $\psi_{i} = 1$, $\forall i = 1, 2, ..., N$ for the CP, and $\psi_{i} \rightarrow \infty$, $\forall i = 1, 2, ..., N$ for the fully RP.

The SIR model is obtained from our model by setting $\eta = 1$, $\delta_1 = \gamma = 0$, $\beta = 0$ and $\alpha = 0$. In this case, the recovery rate is $\delta = \frac{\delta_2}{(1 - \delta_2)}$. Additionally, our framework also includes the SIRS scenario as advanced before. By considering the model introduced in~\cite{Chen2014196} (see also Table~\ref{tab:literature}) the SIRS scheme is recovered by setting $\eta = 1$, $\delta_1 = 0$, $\beta = 0$ and $\alpha = 0$. The system can then be written as a function of $\lambda$, $\delta_2$ and $\gamma$. From equation~\eqref{eq:thr_rp}, such reduction implies that the threshold is given by
\begin{equation}
 \Lambda_{max} \approx \left( \frac{\delta_2}{\lambda (1 - \delta_2)} \right)_c.
\end{equation}
which depends only on $\delta_2$ and $\lambda$.

\subsection{Rumor spreading}

All rumor models in Table~\ref{tab:literature} can be obtained from our general model, except the DK model, since it considers edge sampling and undirected contacts (see section~\ref{sec:review}). The difference between our approach and the previous models is that our approach does not allow an individual to perform two transitions (or two attempts) at the same time step. 

The MK model can be obtained by considering $\eta = 1$, $\beta = 0$, $\delta_1 = \delta_2 = \gamma = 0$ in Eq.~\eqref{eq:rumor_evo}, letting the system be a function of $\alpha$ and $\lambda$. Such model does not have a critical threshold. On the other hand, it is possible to obtain the threshold for the variant introduced by Nekovee et al.~\cite{Nekovee2007}, which considers an spontaneous process to model the lost of interest in the propagation of the rumor. The probability of turning a spreader into a stifler is called $\bar{\delta}$ in the original paper. Such parameter is similar to $\delta_2$ here, which leads us to a threshold equivalent to that in~\cite{Nekovee2007} 
\begin{equation}
 \Lambda_{max} \approx \left( \frac{\delta_2}{\lambda (1 - \delta_2)} \right)_c.
\end{equation}

Finally, the model by Borge-Holthoefer et al.~\cite{Borge-Holthoefer2012} considers a feature that is very common in human communication through online social networks, namely, apathy, in which an ignorant might turn into a stifler after being informed $-$ this is often the case observed in Twitter for instance, where the fact that a tweet appears in a user's timeline does not directly imply that the user spreads it further. This feature has been used to identify influential spreaders in rumor-like dynamics~\cite{Gonzalez2011,Borge-Holthoefer2012,DeDomenico2013, deArruda2014}. The model by Borge-Holthoefer et al.~\cite{Borge-Holthoefer2012} is recovered from our approach by setting $\beta = 0$, $\delta_1 = \delta_2 = \gamma = 0$ in Eq.~\eqref{eq:rumor_evo}. Note that the system's dynamics depends on $\eta$, $\alpha$ and $\lambda$ in this scenario.

\begin{figure*}[t]
\begin{center}
\includegraphics[width=0.95\textwidth]{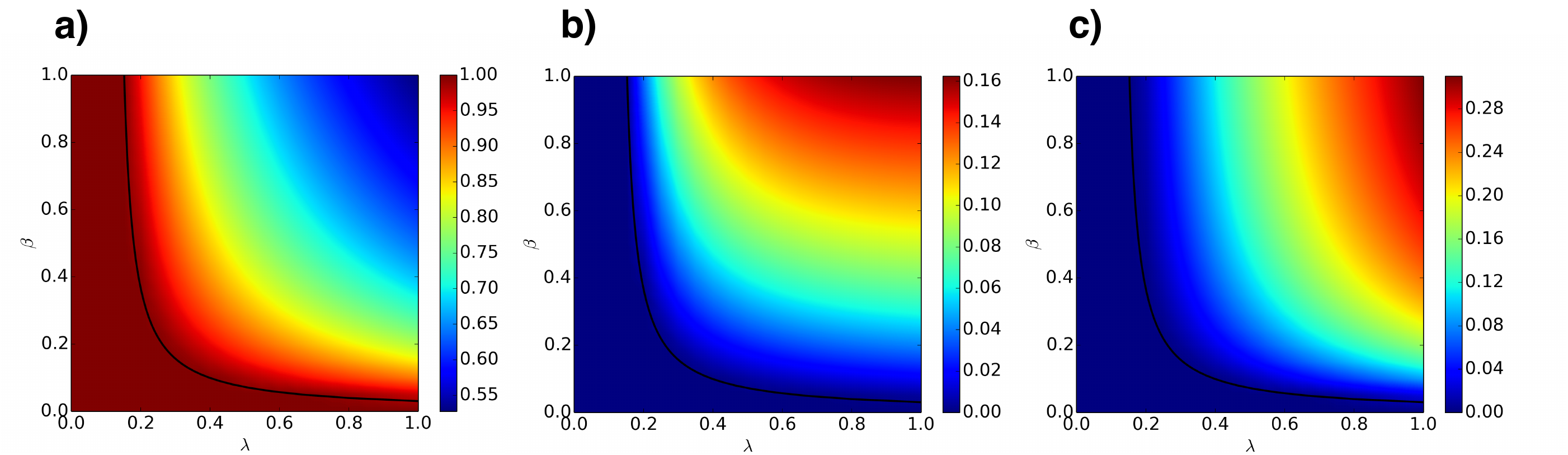}
\caption{$\beta \times \lambda$ phase diagrams at the stationary state for the CP case on a scale-free network with degree distribution $P(k)\sim k^{-\zeta}$ and $\zeta \approx 2.7$. The network is made up of $N = 10^4$ nodes and has an average degree $\langle k \rangle \approx 100$. The rest of dynamical parameters have been set to $\eta = 0.01$, $\delta_1 = \delta_2 = \gamma = 0.1$ and $\alpha = 1.0$. The intensity of the color (as given by the legend) represents the final fraction of ignorants (panel (a)), spreaders (panel (b)) and stiflers (panel (c)). The continuos lines are the analytical values for the critical point.}
\label{Fig:MK_CP}
\end{center}
\end{figure*}

\begin{figure*}[t]
\begin{center}
\includegraphics[width=0.98\textwidth]{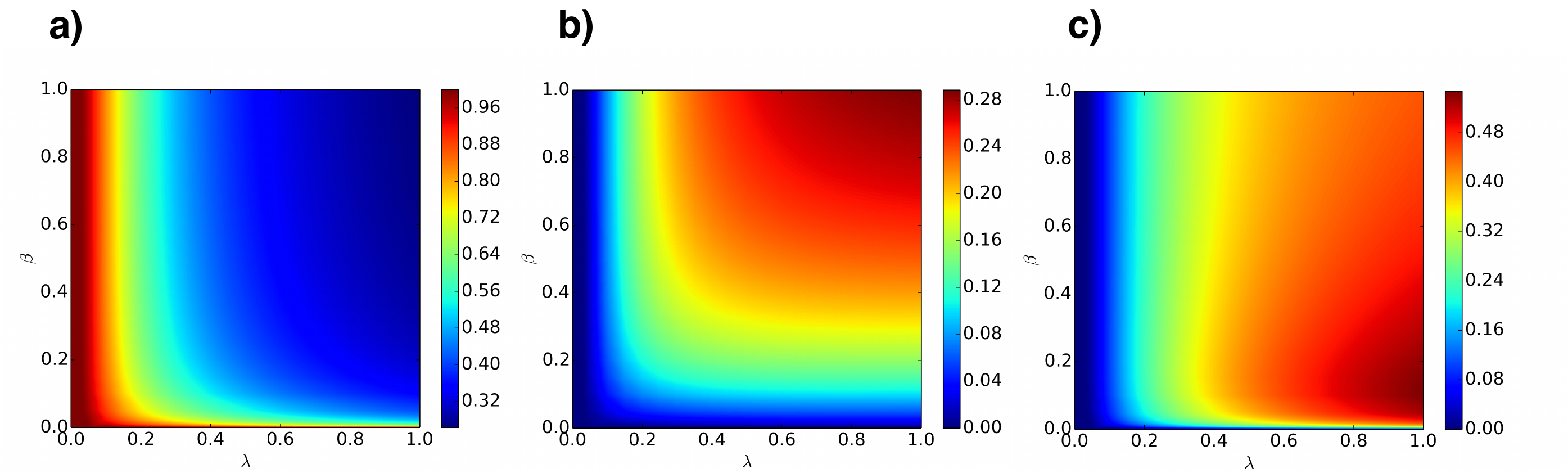}
\caption{$\beta \times \lambda$ phase diagrams as in Fig.~\ref{Fig:MK_CP} but for the RP. The underlying network of contacts has the same features except for the average degree that is $\langle k \rangle \approx 10$. The rest of parameters are $\eta = 0.01$, $\delta_1 = \delta_2 = \gamma = 0.25$ and $\alpha = 1.0$. The intensity of the color (as given by the legend) represents the final fraction of ignorants (panel (a)), spreaders (panel (b)) and stiflers (panel (c)).}
\label{Fig:MK_RP}
\end{center}
\end{figure*}

\section{Numerical analysis and simulations} \label{sec:Numerical}

Once we have got some analytical insights, we next compare results from extensive Monte Carlo (MC) simulations with the numerical solution of the system of equations ~\eqref{eq:rumor_evo} that describes the dynamics of the model, showing that they agree at the micro (e.g., at the individual level) and macro (e.g., at the system level) scales. We obtain the respective phase diagrams for the limiting cases of CP and RP for several combinations of the model's parameters as well as for different topologies of the underlying network of contacts.  

\subsection{Phase diagrams} \label{sec:phase}

First, we present the $\lambda \times \beta$ phase diagrams for the CP and the RP cases. They are obtained by solving the dynamical set of equations ~\eqref{eq:rumor_evo} and varying these parameters from 0 to 1 in intervals of $\Delta\lambda=\Delta\beta=5\times10^{-2}$. In addition, when $\beta < 0.05$ ($\lambda < 0.05$), i.e., near the critical region, we set $\Delta\lambda=\Delta\beta=5\times10^{-3}$. 

Figures~\ref{Fig:MK_CP} and~\ref{Fig:MK_RP} show the $\lambda \times \beta$ phase diagrams obtained from the numerical evaluation of the system of equations~\eqref{eq:rumor_evo} for the CP and the RP, respectively.  We represent (color-coded) in the two figures the fraction of ignorants, spreaders and stiflers in panels (a), (b) and (c), respectively. The underlying network is the same in both cases (scale-free graph with  $\zeta \approx 2.7$, $N = 10^4$), except for the average degree, which is $\langle k \rangle \approx 100$ in Fig.~\ref{Fig:MK_CP} and $\langle k \rangle \approx 10$ in Fig.~\ref{Fig:MK_RP}. The rest of parameters have been set to $\eta = 0.01$, $\delta_1 = \delta_2 = \gamma = 0.1$ and $\alpha = 1.0$ for the CP and to $\eta = 0.01$, $\delta_1 = \delta_2 = \gamma = 0.25$ and $\alpha = 1.0$ for the RP. As discussed in the previous section, it is possible to obtain an analytical expression for the critical point in the CP scenario (Eq.~\eqref{eq:thr_cp}), which is indicated by the continuous black line. This is not possible for the RP.

The comparison of the results shown in Figures~\ref{Fig:MK_CP} and~\ref{Fig:MK_RP} indicates that there are important differences in the system's behavior for the CP and RP schemes, notably with respect to the final fraction of stiflers, which is in most practical scenarios the quantity that we would like to be as higher as possible $-$ as that would mean that the piece of information reached a large fraction of the population and therefore that it was spread efficiently. The CP exhibits a critical point that does not depend on the network structure, and although it cannot be evaluated for the RP in a closed form, the numerical solutions shows that this scenario is more complex with some interesting dependencies on the network structure and the dynamical parameters of the model. As a matter of fact, as it can be seen in Fig.~\ref{Fig:MK_RP}c, there is a nonlinear effect in $\beta$ for fixed values of $\lambda$. Admittedly, for large values of $\lambda\ge 0.7$, when $\beta$ starts to increase from zero, the final density of stiflers also grows. However, at some point (roughly around $\beta\approx 0.1$), this density reaches its maximum value and starts decreasing beyond that value of $\beta$.

\subsection{Monte Carlo simulations} \label{sec:MC}

\begin{figure}[t]
\begin{center} 
\includegraphics[width=0.98\columnwidth]{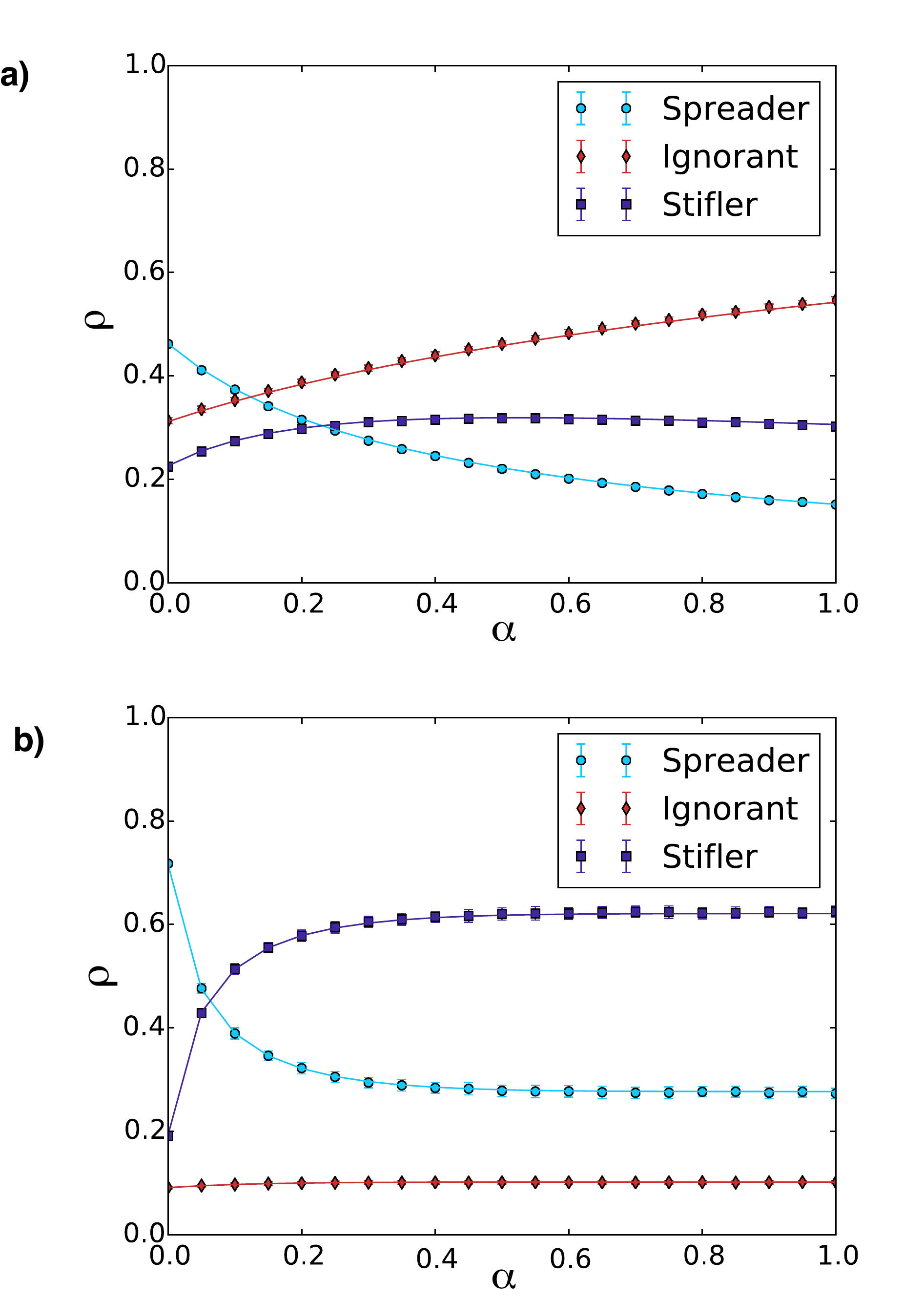}
\caption{Phase diagram at the steady state for (a) a CP considering a scale-free network with $\zeta \approx 2.7$, $N = 10^4$ and $\langle k \rangle \approx 100$, and (b) a RP simulated on top of a scale-free network with $\zeta \approx 2.7$, $N = 10^4$ and $\langle k \rangle \approx 10$. The simulations have been carried out using the following dynamical parameters: $\delta_1 = \delta_2 = \gamma = 0.1$, $\lambda = 1$, $\beta = 0.5$ and $\eta = 1$. The continuous lines are the theoretical predictions, whereas the symbols are the results of MC simulations. The standard deviation has approximately the size of the symbols. }
\label{Fig:MK_Comp}
\end{center}
\end{figure}

In order to check whether our analytical and numerical solutions are accurate at describing both individual states and the macroscopic behavior of the system, we have performed large-scale Monte Carlo simulations. All results reported henceforth are averages taken over at least 100 MC simulations with an initial fraction of spreaders and ignorants equal to $\rho^Y(0) = 0.01$ and $\rho^X(0) = 0.99$, respectively.

\begin{figure}[t]
\begin{center}
\includegraphics[width=0.98\columnwidth]{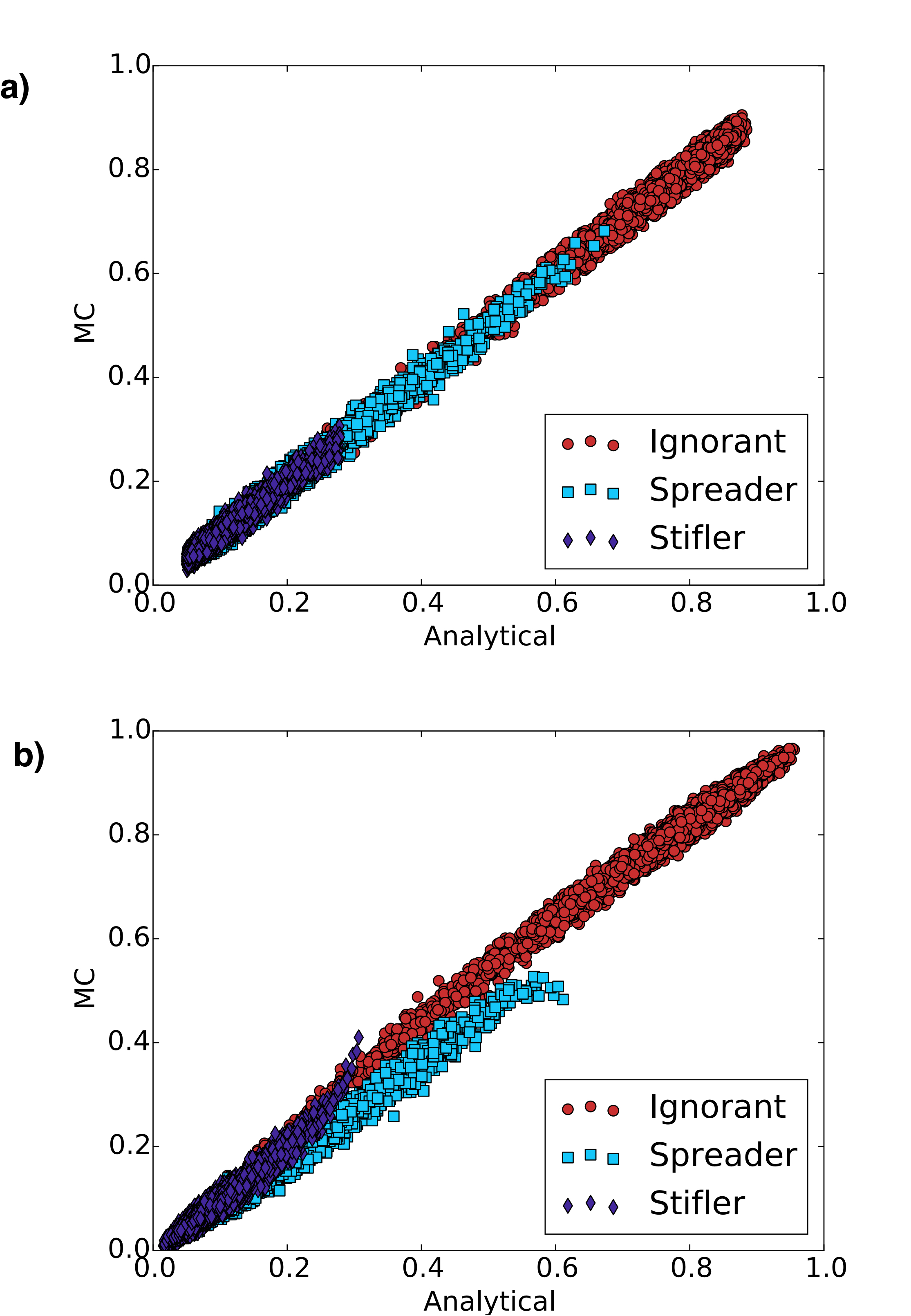}
\caption{Probability of finding a node $i$ in any of the dynamical states of the model ($X$, $Y$ or $Z$) obtained by solving analytically the system of equations~\eqref{eq:rumor_evo} as compared with Monte Carlo simulations. The probabilities are calculated by averaging $10^3$ simulations. Panel (a) corresponds to the CP, whereas the RP is represented in panel (b).}
\label{Fig:Micro}
\end{center}
\end{figure}

In Fig.~\ref{Fig:MK_Comp}, we present results for the dependency of the densities of $X$, $Y$ and $Z$ as a function of $\alpha$ with all the other model parameters fixed. The $\rho \times \alpha$ plots correspond to the CP (panel (a)) and the RP (panel (b)) scenarios. As it can be seen, the continuous lines, which are obtained by numerically evaluating Eq.~\eqref{eq:rumor_evo} perfectly match results from MC simulations in both limiting cases. In order to check whether this agreement is also verified at the individual scale, we have represented in a scatter plot, see Fig.~\ref{Fig:Micro}, the probability of finding an individual in the ignorant, spreader or stifler class at the stationary state. Despite of finite size effects and stochastic fluctuations, the analytical results capture the micro dynamical states in the large $t$ limit, since there is a strong correlation between the probabilities obtained from MC and the solutions of the system Eq.\eqref{eq:rumor_evo}. These results thus convincingly show that one can explore the model in terms of either the individual probabilities or the macroscopic expected values by solving the equations describing the system's dynamics, without the need to rely on extensive and costly MC simulations. This is, from a practical point of view, an added value of the proposed framework, as we shall discuss later on.

\begin{figure*}[!t]
\begin{center}
\includegraphics[width=0.95\textwidth]{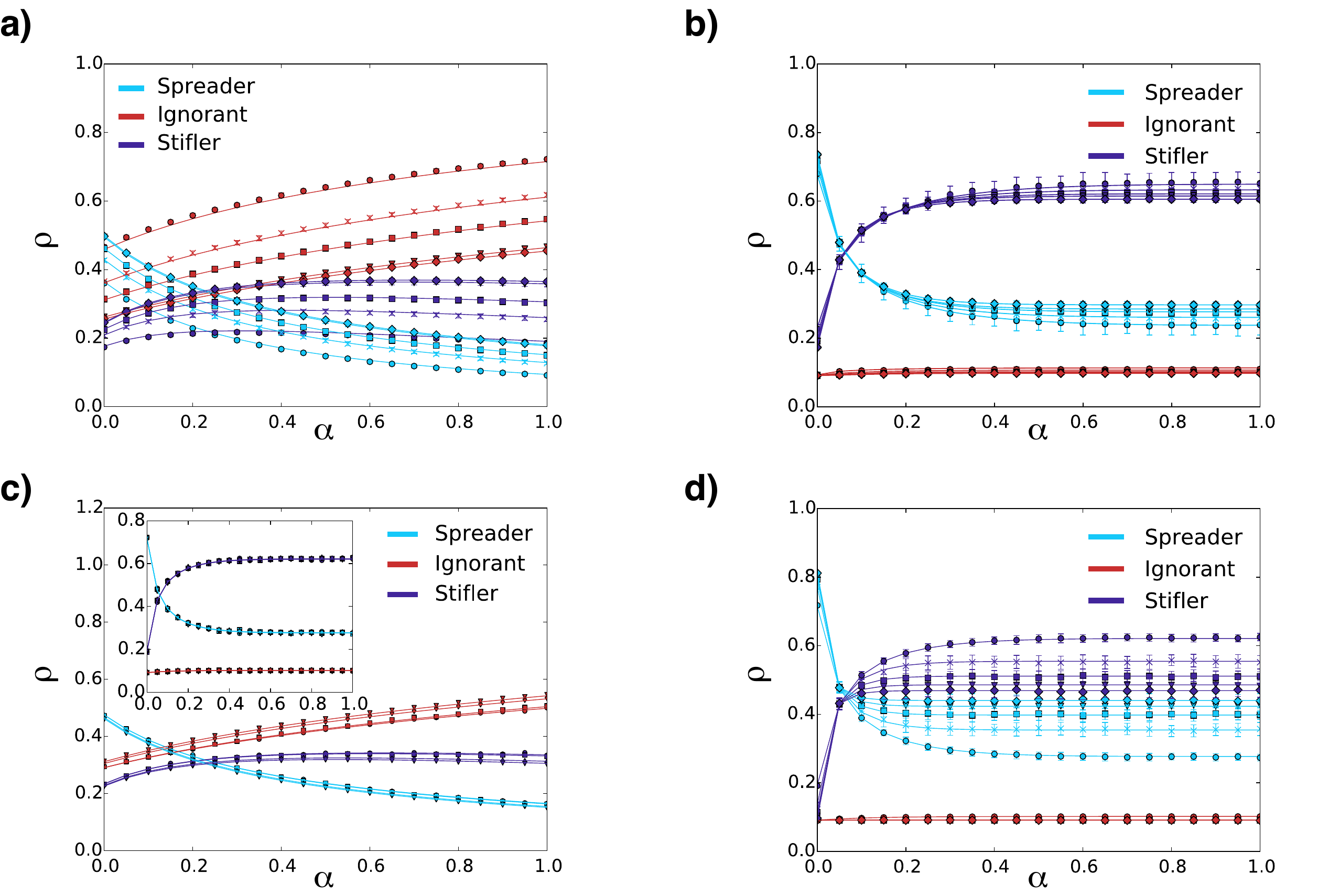}
\caption{Phase diagram of the system's dynamics at the steady state as a function of $\alpha$. In all panels the rest of parameters have been fixed to $\delta_1 = \delta_2 = \gamma = 0.1$, $\lambda = 1$, $\beta = 0.5$ and $\eta = 1$ and the continuous lines are the theoretical predictions, while the symbols correspond to the MC simulations. Panels (a) and (b) show results obtained varying the exponent $\zeta$ of the power law degree distribution, $P(k) \sim k^{-\zeta}$, of the underlying networks of contacts of size is $N = 10^4$ and $\langle k \rangle \approx 100$ (CP, panel (a)) and $\langle k \rangle \approx 10$ (RP, panel (b)). We considered the following exponents $\zeta$: $2.3$ ($\bullet$), $2.5$ ($\times$), $2.7$ ($\square$), $3.3$ ($\triangledown$) and $3.5$ ($\diamond$). The effects of the network size are shown in Panel (c) for scale-free networks ($P(k) \sim k^{-\zeta}$ with $\zeta \approx 2.7$) with $\langle k \rangle \approx 100$ (main plot) and $\langle k \rangle \approx 10$ (inset) and the following sizes: $N = 5 \times 10^2$ ($\bullet$), $N = 10^3$ ($\square$), $N = 5 \times 10^3$ ($\diamond$) and $N = 10^4$ ($\triangledown$). Finally, in panel (d), we represent results obtained for the RP and different average degrees: $\langle k \rangle \approx 10$ ($\bullet$), $\langle k \rangle \approx 20$ ($\times$), $\langle k \rangle \approx 35$ ($\square$), $\langle k \rangle \approx 45$ ($\triangledown$) and $\langle k \rangle \approx 60$ ($\diamond$). The rest of network's parameters are the same as in panel (c).}
\label{Fig:Gamma}
\end{center}
\end{figure*}

Spreading processes have been shown to be greatly affected by the topological features of the networks on top of which they take place. It is therefore of further interest to investigate network effects on the dynamical evolution of the spreading phenomenon. To this end, we have explored the impact of three characteristics of the contact networks, namely, the exponent of the degree distribution $P(k)$ for scale-free graphs $-$ for which $P(k)\sim k^{-\zeta}$$-$, the size of these networks and finally how dense they are by tuning the average degree of the network's nodes. Figure~\ref{Fig:Gamma} shows the results of these analyses. First, we note that in all cases, the previous agreement between the numerical solution of the system's equations and MC simulations still holds. Other aspects worth highlighting include the fact that the value of $\zeta$ influences the outcomes of the spreading process mainly for the CP, unless $\zeta$ is larger than $3$. This is because for larger exponents, the networks are effectively equivalent to homogeneous graphs, and therefore there are no hubs anymore, leading to fairly similar results independently of the specific value of $\zeta$. Another interesting effect of the network structure, and in particular of $\zeta$, is the one observed in Fig.~\ref{Fig:Gamma}b. Admittedly, the final density of ignorants is independent of this network parameter. In other words, variations of the exponent mainly affect the number of spreaders and stiflers, with the number of ignorants remaining roughly constant. 

As for finite size effects, our results, see Fig.~\ref{Fig:Gamma}c, show that they do not appear to play a major role on the rumor spreading dynamics. This is not the case with respect to the last parameter analyzed, the average degree of the network. As it can be seen in Fig.~\ref{Fig:Gamma}d, the steady-state fractions of spreaders and stiflers strongly depend on the density of connections, while the fraction of ignorants is (as it happened with respect to $\zeta$) almost the same for different average degrees. In summary, it seems that the most important dependencies with respect to the networks' topological properties are given by its heterogeneity (characterized by $\zeta$) and the density of connections (e.g., $\langle k \rangle$). Both features mainly affect the ratio between spreaders and stiflers at the stationary state, but do not significantly affect the final number of ignorants. Altogether, and from a practical point of view, one can then conclude that, with the exception of the heterogeneity of the degree distribution, the size and average degree of the networks have a somewhat minimal impact on the propagation of the rumor, at least in what concerns its main outcome: the number of individuals that learned the rumor $-$both spreaders and stiflers represent states in which individuals are aware of the news, that is, they have been reached by the contagion process (but also see below). 

\begin{figure}[t]
\begin{center}
\includegraphics[width=0.95\columnwidth]{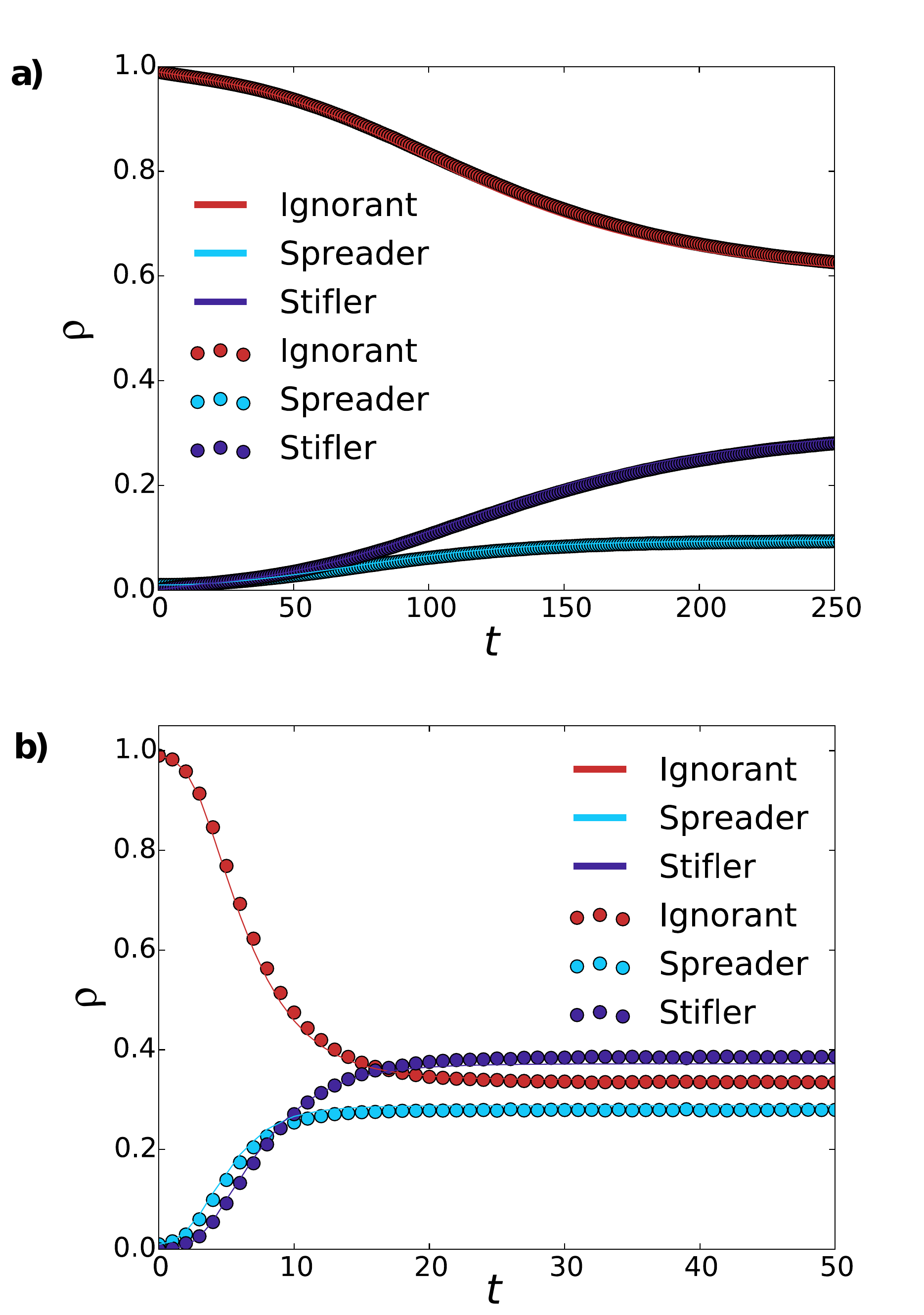}
\caption{Time evolution of the average probabilities for the (a) CP and (b) RP. The continuous lines are the theoretical predictions, while the symbols are obtained by Monte Carlo simulations.}
\label{Fig:Time}
\end{center}
\end{figure}

Finally, we also show Fig.~\ref{Fig:Time} that the numerical solution of the system~\eqref{eq:rumor_evo} and Monte Carlo simulations are in good agreement not only in the large $t$ limit, but also when we explore the system's behavior at intermediate times, i.e., in transient states. The figure shows the evolution in time of the fraction of ignorants, spreaders and stiflers for both the CP and RP limiting cases when the substrate network is a scale-free graph with $\zeta \approx 2.7$, $N = 10^4$ and $\langle k \rangle \approx 100$. As we will see in the next section, the fact that the temporal dynamics can be captured accurately by numerically solving the set of equations ~\eqref{eq:rumor_evo} could be used to study real spreading dynamics for which highly resolved temporal data is available. 

\begin{figure}[t]
\begin{center}
\includegraphics[width=0.95\columnwidth]{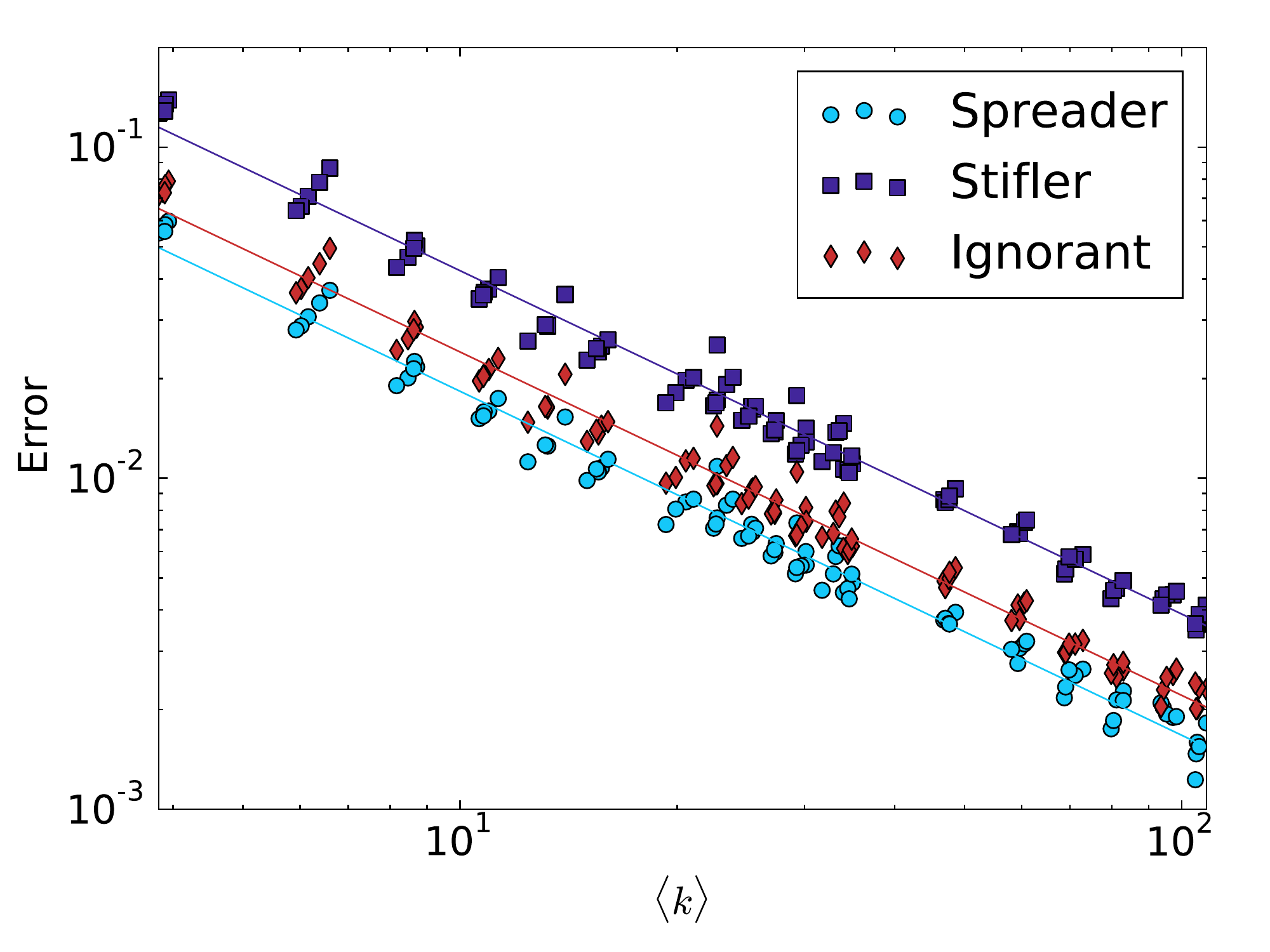}
\caption{Phase diagram at the steady state for the CP considering a scale-free network with $\zeta \approx 2.7$, $N = 10^4$ and the dynamical parameters $\delta_1 = \delta_2 = \gamma = 0.1$ and $\lambda = 1$. The continuous lines are obtained by the least squares method.}
\label{Fig:Error_CP_k}
\end{center}
\end{figure}

We round off this section by pointing out some limitations of the mean field approximation. Gleeson et al.~\cite{Gleeson2012} discussed that the accuracy of the mean field theory is higher when three main assumptions are satisfied, i.e., (i) vanishing local clustering, (ii) non-modular network organization and (iii) absence of dynamical correlations. Conditions (i) and (ii) depend on the structure of the network. They are satisfied in our analysis, because we consider the configuration model, which satisfies the properties (i) and (ii). We will next analyze how our model performs in real online networks. However, the dynamical correlations might represent a major source of error, since we assume that the state of each node is independent of the state of the rest of vertices. Nevertheless, a node is more likely to be informed from one of its neighbors than from other vertices. In~\cite{Gleeson2012} the authors exemplified such effect by considering a SIS dynamics, suggesting that the error is reduced with the increase of the mean first neighbor degree. To quantify the error between our numerical experiments and MC simulations, we set the dynamical parameters to $\delta_1 = \delta_2 = \gamma = 0.1$, $\lambda = 1$ and vary $\alpha$. The differences between the theoretical predictions and MC simulations are then quantified by the absolute error. We found (results not shown) that, for the RP, the error vanishes as soon as $\langle k \rangle \geq 10$, whereas for the case of a CP, the error is close to zero only for $\langle k \rangle \geq 100$, as shown in Fig.~\ref{Fig:Error_CP_k}.

\section{Applications to real social systems} \label{sec:social}

\begin{figure*}[t]
\begin{center}
\includegraphics[width=0.95\textwidth]{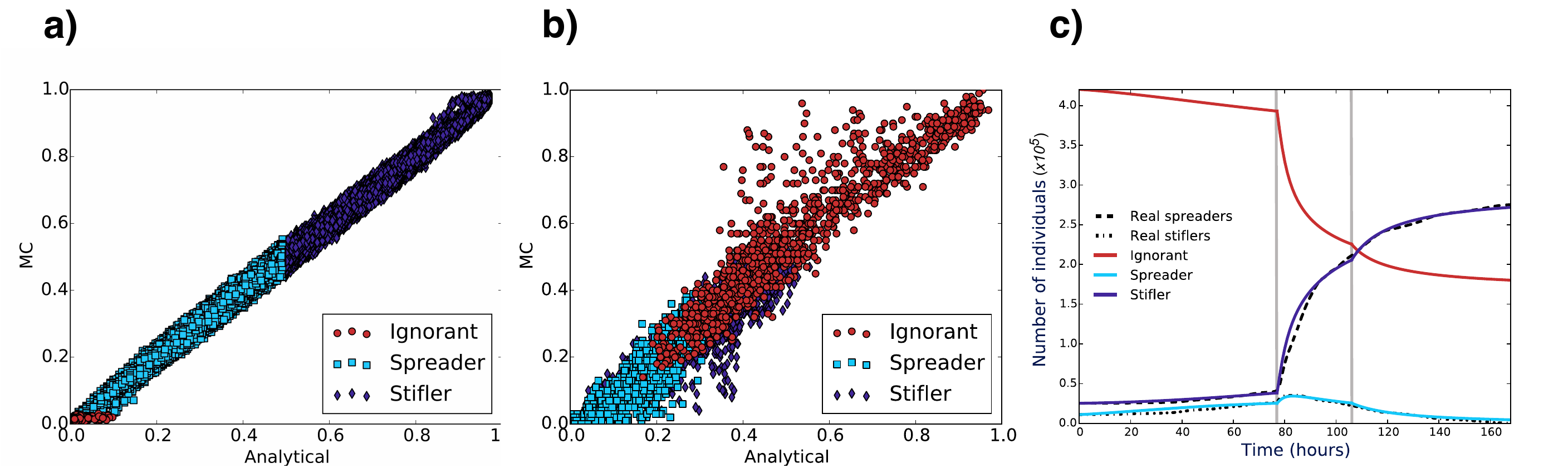}
\caption{Application of the proposed discrete time Markov chain formulation to the analysis of spreading processes that take place on top of social networks. Panels (a) and (b) show the probability of node $i$ belonging to state $X$, $Y$ or $Z$ obtained by solving numerically the system of equations~\eqref{eq:rumor_evo}) and by Monte Carlo simulations. Panel (a) corresponds to a RP that takes place over the Twitter network, where as panel (b) has been obtained simulating a CP on the email network. The model parameters: $\lambda = 1.0$, $\eta = 0.01$, $\delta_1 = \delta_2 = \gamma = 0.01$, $\beta = 1.0$ and $\alpha = 1.0$ for the RP; and $\lambda = 1.0$, $\eta = 0.01$, $\delta_1 = \delta_2 = \gamma = 0.1$, $\beta = 1.0$ and $\alpha = 1.0$. Results for MC simulations are averages over $10^3$ different simulations. Panel (c) depicts the time evolution of the Higgs boson rumor spreading, taking into account tweets of the dataset shown in~\cite{DeDomenico2013}. The dashed lines represent the real data, while the continuous lines are the numerical solutions of our model. The vertical lines mark the three time windows used, with the following parameters: (i) $0 \leq t \leq 77$, $\lambda = 0.00025$, $\alpha = 0.0002$, $\delta_2 = 0.0001$ and $\eta = 0.85$; (ii) $77 < t \leq 106$, $\lambda = 0.021$, $\alpha = 0.00075$, $\delta_2 = 0.0015$ and $\eta = 0.17$; (iii) $t > 106$, $\lambda = 0.065$, $\alpha = 0.002$, $\delta_2 = 0.002$ and $\eta = 0.01$.}
\label{Fig:Real_micro}
\end{center} 
\end{figure*}

Up to now, we have explored the model in computer-generated networks. While these networks share some of the topological features that have been found in real-world systems, they don't account for all of them, specially when it comes to clustering properties or several kinds of correlations. It is therefore important to run our model on top of real systems and check whether the results reported so far also hold for more realistic topologies. In doing so, we will also show how the model could be used to get a better understanding of the mechanisms driving real spreading phenomena. In what follows, we inspect whether the model gives accurate results at the microscopic scale for both the CP and the RP scenarios, as we have done before for the synthetic networks. Finally, we also reproduce the temporal dynamics of a real spreading process that took place over online social networks (Twitter in our analysis) when the confirmation of the existence of the Higgs boson was announced. 

To verify that our model performs well also in real networks, we use two social networks. The first is an email network, in which messages are mostly spread as a contact processes (that is, one-to-one communication at a time), and the second network considered is the contact patterns of Twitter. The latter is an online social system where the information dynamics is mainly of the form one-to-all, that is, a user posts a message that reaches out all the user's followers at the same time, thus corresponding to the RP limit in our formulation. We also consider $\eta = 0.01$ to simulate the apathy as in~\cite{Borge-Holthoefer2012}.

The email contact network was created from emails exchanged between users within the Universitat Rovira i Virgili~\cite{PhysRevE.68.065103}. The network is composed by $N = 1133$ nodes and $\langle k\rangle = 9.6$; connections are directed and unweighted. However, here we consider an undirected version of this network. The Twitter network was extracted from the mobilizations in Spain during 2011~\cite{Borge-Holthoefer2011, Gonzalez2011, Borge-Holthoefer2012b}. Here we consider a simplified version of this network, composed by $N = 85712$ nodes and $\langle k\rangle = 109.9$, with undirected and unweighted contacts. Both networks have an average clustering coefficient $\langle cc \rangle \approx 0.22$. In addition, while the email network is slightly assortative, $r = 0.08$, the Twitter network is disassortative $r = -0.14$.

We compare the results obtained via MC simulations of the model with the numerical solution of Eq.~\eqref{eq:rumor_evo} in Fig.~\ref{Fig:Real_micro}a and b using the Twitter and the email networks as underlying graphs, respectively. The results show that despite the new topological features of the real networks, the agreement is still very good. Panel (a) corresponds to a RP and the probabilities of finding a node in each of the three possible states at the large time limit match fairly well. This indicates that the new topological ingredients that were not present in the synthetic networks play a minor role when it comes to evaluate the accuracy of the discrete Markov chain formulation $-$ or at the very least for this particular network. Panel (b) on the contrary shows the results obtained when simulating a CP on top of the email network. Again, the numerical solutions of the set of equations describing the dynamics of the system agree well with MC simulations, albeit having larger deviations and more dispersion along the diagonal line. This could be due to the fact that $\langle k\rangle = 9.6$ for the email network, and as discussed and shown earlier (see Fig.~\ref{Fig:Error_CP_k}), for a CP errors are vanishingly small only beyond $\langle k\rangle \approx 100$. Other new  topological characteristics or even some dynamical correlations~\cite{Gleeson2012} might have an impact as well. Additionally, note that another possible source could be the relative small size ($N=1133$) of the email network. Admittedly, the Twitter network also has triangles, modular organization and degree-degree correlations~\cite{Borge-Holthoefer2012, Borge-Holthoefer2011}, but it is an order of magnitude larger. 

We also apply the formalism to simulate the time evolution of a real information dissemination process. Specifically, we have modeled the temporal dynamics of the rumor spreading on Twitter during and after the announcement of the discovery of a new particle with the features of the elusive Higgs boson on 4th July 2012. Such dataset was formerly analyzed in~\cite{DeDomenico2013}. Here, we consider the giant component of an undirected network of the friend/follow network, which  is composed by $N = 456626$ individuals, with $\langle k \rangle = 54.79$, an average clustering coefficient $\langle cc \rangle = 0.189$ and assortativity $r = -0.098$. This database describes the timestamps (in seconds) of mentions, replies and retweets. In our analysis we consider these three timestamps as events. However, if two events occur at the same time and are generated by the same user, we consider them as just one event. For instance, if an user retweets a mention we would have two events at the same time, one for the mention and the other for the retweet. In such a case we consider it as just one tweet.

Moreover, how to assign the state a node belongs to (i.e., spreader or stifler) is not trivial, because we cannot distinguish between: (i) a spreader turns into a stifler, then recovers the interest in the rumor and becomes a spreader again or (ii) a spreader that remains as a spreader during the observation time. Here, we assume that an individual is an ignorant if he/she did not tweet about the Higgs boson. An ignorant becomes a spreader the time he/she first tweets about this topic, and remains as such up to its last tweet, when it becomes a stifler. The only special case considered are the users that tweeted just once. In such a case, we consider that these individuals are spreaders during just one time step, i.e., 1 unit time, and they become stiflers the next time-stamp. Additionally, the initial conditions are given by the activities before the observation time window, which implies to start with a certain fraction of spreaders and also stiflers, since some users that tweeted about the Higgs boson do not tweet during the observation time window. Note that the modeling assumes that a user is aware of the rumor only if he/she tweets about it. Finally, it is important to define $\Delta t$. For our analysis, we have assumed that each discrete time window represents 1 hour of the real data. Such a choice implies that we are not able to distinguish events that occur at a faster rate. However, setting $\Delta t=1h$ drastically reduces the computational cost. We remark that our goal here is to verify whether our model is capable of describing the behavior observed in the real data, not to perform a detailed forensic analysis of the actual rumor spreading process. 

In order to fit simulations of the model with the data, we assume that the forgetting mechanisms can be neglected, since the total observation time is reasonable small (1 week), implying that $\delta_1 = \gamma = 0$. Moreover, we assume $\beta = 0$, since we consider that there is no transition from stifler to spreader, i.e., the possibility of recovering the interest in the rumor is neglected $-$ this is in part also due to the constraint that we can't distinguish such transitions in real data. Additionally, observe that $\delta_2$ is the probability that a spreader loses the interest about the rumor spontaneously, while $\alpha$ represents the probability of turning into stiflers after contacting spreaders or stiflers. Note that $\eta$, $\gamma$ and $\alpha$ might produce similar effects, depending on the defined time-steps. 

Figure~\ref{Fig:Real_micro}c compares the time evolution of the spreading dynamics as extracted from the real data with results obtained from our model using three different time windows. The latter is needed as it is known that in viral processes like the one we are analyzing, there are different phases of the dynamics: an early stage in which the number of messages exchanged increases slowly (subcritical regime) followed by an explosive period (critical and supercritical regimes) that signals the moment beyond which the piece of news goes viral and finally a phase in which the active spreaders die out and stop propagating the rumor any farther~\cite{Borge-Holthoefer2012}. For the first time window, we thus assign initial probabilities of being in state $X$, $Y$ or $Z$ as given by the real data, but considering them to be the same for every node. For the following time windows, these probabilities come from the output of the previous simulation window. On the first time window, $0 \leq t \leq 77$, we consider $\lambda = 0.00025$, $\alpha = 0.0002$, $\delta_2 = 0.0001$ and $\eta = 0.85$. On the second, $77 < t \leq 106$, $\lambda = 0.021$, $\alpha = 0.00075$, $\delta_2 = 0.0015$ and $\eta = 0.17$. Finally, for $t > 106$, $\lambda = 0.065$, $\alpha = 0.002$, $\delta_2 = 0.002$ and $\eta = 0.01$. 

The results show that the model is indeed able to accurately reproduce the temporal evolution of the real spreading dynamics. We however stress that we chose those sets of parameters by simple inspection, i.e., we did not apply any fitting algorithm. In order to obtain a better fit, one may use statistical inference tools or even a simulated annealing algorithm. As this is beyond the scope of this work, we leave this line of research as a potential future work. The simulation performed is nevertheless worth carrying out. The fact that a model like the one discussed here could be adapted without the use of sophisticated fitting algorithms to describe the temporal dynamics of a real rumor spreading process is an important step towards getting a better understanding of the mechanisms at work in this real contagion dynamics. Admittedly, without knowing the parameters, real-time projections of the temporal evolution of the rumor dynamics can not be done. But once we know the parameters that give the best matching with the real data, we are in the position to know which mechanisms (that is, what transitions) are more important and which do not, thus gaining valuable phenomenological insights. In turn, given the universal features behind spreading processes, this would allow to perform other analyses, such as detecting who are potential candidates to be influential spreaders in other online social systems or for the propagation of other rumors in the very same network.

\section{Conclusions} \label{Sec:conclusion}

Spreading processes play an important role in nature, society and engineering~\cite{Satorras015}. Due to their relevance, several models have been developed aiming at understanding, modeling and predicting how viruses, diseases, rumors and information propagate through complex networked systems. Despite the theoretical approaches that have been developed to model, for instance, epidemic and rumor dynamics~\cite{Satorras015}, there are still important mechanisms that have not been taken into account, nor there exists a coherent and unifying theoretical and computational framework that deals with as many of these mechanisms as possible at the same time. In this paper, we provide such a unifying methodology using a discrete Markov chain approach, which includes and generalizes previous epidemic and rumor models by accounting for non-traditional behaviors, such as apathy, forgetting, and lost and recovering of interest.

We have focused our study on the theoretical and numerical analysis of the model, and have shown that results obtained by numerically solving the system of equations describing the system's dynamics are in good agreement with extensive Monte Carlo simulations for three different scales: macro, micro and temporal. Regarding analytical results, we have obtained closed forms for the thresholds and the steady state densities of individuals in the different dynamical classes for the CP scenario and some special cases of the RP. Additionally, we have throughly analyzed the influence of the model parameters as well as several network properties on the spreading dynamics. Our findings indicate that using synthetic networks could help getting a first insight into what are the effects of the different mechanisms at work with a high degree of accuracy. Finally, we have studied the propagation of rumors considering real networks on top of which both a contact process and a reactive process might take place. Importantly, we were able to reproduce the time evolution of a rumor propagation using Twitter's user activity during and after the announcement of the discovery of the Higgs boson.

Our formalism is general and covers many models on the literature. This opens new opportunities for the analysis of real data. It is also worth stressing that in information spreading on real systems like online social networks, rumor models play an important role. With this work, we have provided a framework that paves the way to developing new algorithms that could explore accurately and very fast different mechanisms and scenarios for viral information spreading. At variance with disease spreading, in which one is constrained to model a real outbreak, when it comes to design new ways to efficiently disseminate information, one is free to design the mechanisms that would optimize such a spreading process. In other words, one can design the viral process without being constrained to fit a given past or ongoing outbreak. To this end, performing Monte Carlo simulations would be prohibitively costly given the size of the parameters' phase space. This practical hurdle might be surmounted by using the discrete Markov chain approach proposed here, as the computational cost of solving the set of equations is significantly low as compared to Monte Carlo simulations. This is of special relevance when we are dealing with online social systems, whose sizes go from a few hundreds of individuals to millions of users. We plan to explore this line of research in the future. 

\acknowledgments
F. A. R acknowledge CNPq (grant 305940/2010-4) and Fapesp (grant 2013/26416-9) for financial support. G. F. A acknowledges Fapesp for the sponsorship provided (grants 2012/25219-2 and 2015/07463-1). P. M. R acknowledges Fapesp 2015/03868-7. E.C is funded by the EC FET-Proactive Project Multiplex (grant 317532). Y. M. acknowledges financial support from the Government of Aragon (grant to the group FENOL),  MINECO (Spain) through grant FIS2011-25167 and by the EC through the FET-Proactive Project Multiplex (grant 317532).


\end{document}